\documentclass[]{jfm}
\usepackage{graphicx}
\usepackage{newtxtext}
\usepackage{newtxmath}
\usepackage{natbib}
\usepackage{hyperref}
\usepackage{float}
\usepackage{subfig}
\usepackage{dcolumn}
\usepackage{bm}
\usepackage{gensymb}
\usepackage{tabularx}
\usepackage{color}
\usepackage{braket}
\usepackage{comment}
\hypersetup{colorlinks = true,
	    urlcolor   = blue,
	      citecolor  = blue,}

\title{Natural convection in a vertical channel. Part 2. Oblique solutions and global bifurcations in a spanwise-extended domain}
\shorttitle{Natural convection in a vertical channel. Part 2}
\shortauthor{Z. Zheng, L.S. Tuckerman and T.M. Schneider}

\author{Zheng Zheng\aff{1},
        Laurette S. Tuckerman\aff{2} \corresp{\email{laurette.tuckerman@espci.fr}}
        \and Tobias M. Schneider\aff{1}}

\affiliation{
    \aff{1} Emergent Complexity in Physical Systems Laboratory (ECPS), \'Ecole Polytechnique F\'ed\'erale de Lausanne, CH 1015 Lausanne, Switzerland
    \aff{2} Physique et M\'ecanique des Milieux H\'et\'erog\`enes (PMMH), CNRS, ESPCI Paris, PSL University, Sorbonne Universit\'e, Universit\'e de Paris, 75005 Paris, France}

\begin{document}
\maketitle

\begin{abstract}
Vertical thermal convection is a non-equilibrium system in which both buoyancy and shear forces play a role in driving the convective flow. Beyond the onset of convection, the driven dissipative system exhibits chaotic dynamics and turbulence. In a three-dimensional domain extended in both the vertical and the transverse dimensions, \cite{Gao2018} have observed a variety of convection patterns which are not described by linear stability analysis. We investigate the fully non-linear dynamics of vertical convection using a dynamical-systems approach based on the Oberbeck–Boussinesq equations. We compute the invariant solutions of these equations and the bifurcations that are responsible for the creation and termination of various branches. We map out a sequence of local bifurcations from the laminar base state, including simultaneous bifurcations involving patterned steady states with different symmetries. This atypical phenomenon of multiple branches simultaneously bifurcating from a single parent branch is explained by the role of $D_4$ symmetry. In addition, two global bifurcations are identified: first, a homoclinic cycle from modulated transverse rolls and second, a heteroclinic cycle linking two symmetry-related diamond-roll patterns. These are confirmed by phase space projections as well as the functional form of the divergence of the period close to the bifurcation points. The heteroclinic orbit is shown to be robust and to result from a 1:2 mode interaction. The intricacy of this bifurcation diagram highlights the essential role played by dynamical systems theory and computation in hydrodynamic configurations.
\end{abstract}


\section{Introduction}
\par Vertical convection, in which a layer of fluid is confined between two vertical plates maintained at different temperatures, is relevant for industrial applications, including the control of insulation properties of double-glazed windows. Vertical convection also serves as a model system in the geophysical context to describe convectively driven flows in the earth, the ocean and the atmosphere. Moreover, vertical convection is a fundamental hydrodynamics problem in its own right, as a prototype for studying pattern formation mechanisms within spatially extended driven dissipative nonlinear out-of-equilibrium systems. In our companion paper \cite{Zheng2023part1}, we studied a domain in which the transverse (or spanwise) direction was taken to be of the same length as the distance between the plates (the wall-normal direction), with the vertical dimension (parallel to gravity) chosen large compared to both. Consequently, flow patterns are primarily two-dimensional (2D), with variations predominantly in the vertical and wall-normal direction. Here, we will consider an extended three-dimensional (3D) geometry, in which the transverse and vertical dimensions are both large compared to the inter-plate spacing and thus flow patterns vary in two extended directions.  

\par We begin by briefly surveying 3D numerical investigations of vertical convection. \cite{Chait1989, henry1998two, xin2002extended} analyzed the instability of 2D nonlinear flow (transverse rolls) to 3D perturbations in order to determine when and whether the flow could be assumed to be 2D. In Rayleigh-B\'enard convection, the stability thresholds in Rayleigh number ($Ra$), Prandtl number ($Pr$), and 2D roll wavelength delimit a volume that is called the Busse balloon \citep{busse1978non}, named after the researcher who has been at the forefront of pattern formation research in Rayleigh-B\'enard convection. Busse later also transferred his analysis to vertical convection. Using the approximation (corresponding to infinite thermal diffusivity) that the temperature retains its linear conductive profile, \cite{nagata1983three} computed a fully nonlinear 3D solution which is probably the diamond roll state (FP2) to be described in section \ref{part2_Equilibria}. Such 3D solutions have sometimes been termed tertiary solutions, while the laminar and 2D transverse roll solutions are called primary and secondary, respectively. \cite{Clever1995} extended the computation of 3D solutions to $Pr=0.71$, corresponding to convection in air, the case we study in this paper. 

\par \citet{Gao2013, Gao2015, Gao2018} combined linear and weakly non-linear theory as well as direct numerical simulations to study the three-dimensional flow. \citet{Gao2013,Gao2015} studied the equilibria and periodic orbits in a computational domain of size $[L_x, L_y, L_z] = [1, 1, 10]$, the same domain we consider in our companion paper \citet{Zheng2023part1}. In order to study secondary instabilities in the transverse direction of the 2D steady rolls, \cite{Gao2018} computed their linear stability. Their analysis showed two types of instabilities, with spanwise wavelengths of about four and eight. They consequently extended the spanwise length of the domain from unity to $L_y=8$ to capture both instabilities. In addition, when $L_z=9$ the Rayleigh number thresholds of both types of 3D instabilities are close, motivating them to decrease $L_z$ from $10$ to $9$ in order to study the competition between both instabilities destabilizing 2D rolls. Like a spanwise domain size of $L_z=10$, a domain with $L_z = 9$ also accommodates four co-rotating rolls in the primary instability of the base state and is large enough to allow interactions between rolls. \cite{cimarella2017routes} and \cite{cingi2021direct} unsuccessfully attempted to explain the results of \cite{Gao2013, Gao2018} from a bifurcation-theoretic point of view. 

\par In this paper, we study vertical convection in air ($Pr=0.71$) in the configuration $[L_x,L_y,L_z]=[1,8,9]$. Similarly to the approach described in \citet{Zheng2023part1}, we extend previous studies by \citet{Gao2018} that were based primarily on time-stepping by using numerical continuation and stability analysis. This unravels the bifurcation-theoretic origins of complex flows and the connections between them. This approach of explaining patterns and their dynamics in terms of equilibria and periodic orbits has been successfully applied to inclined layer convection where fascinating convection patterns were previously observed in direct numerical simulations and experimentally by \cite{Daniels2000}. Through a numerical bifurcation analysis, \citet{Reetz2020a} and \citet{Reetz2020b} identified the invariant solutions underlying most of the patterns and constructed bifurcation diagrams connecting them. These invariant solutions capture key features and dynamics of the observed patterns and the bifurcation diagrams reveal their origin. Here, we follow the same strategy to explain flow patterns in vertical convection in a somewhat larger domain.

\par Using parametric continuation techniques that can follow states irrespective of their stability, we will describe the discovery of three new branches of steady states, which, together with those observed by \cite{Gao2018} via time integration, brings the number of branches observed thus far to six. Several of these new states bifurcate simultaneously, at the same value of the control parameter, despite not being related by symmetry. We have shown that this otherwise non-generic phenomenon is explained by the fact that the parent branches have $D_4$ symmetry; see \citet{swift1985bifurcation, Knobloch1986, Chossat1994, bergeon2001three, Reetz2020b}. In our geometry, $D_4$ symmetry leads to simultaneous bifurcations to states that are aligned with respect to the transverse and vertical directions, and others which are diagonal with respect to them. Competition between aligned and diagonal states is also seen in two periodic orbits (observed by \cite{Gao2018}), that consist of diagonal excursions from more aligned states. We have also discovered two new periodic orbits.

\par Most of the steady states and periodic orbits that we have identified are unstable. While these are not directly observed in time-dependent simulations, following unstable branches is essential for understanding the origin of stable states and for constructing a bifurcation diagram unifying the solutions to a problem. Moreover, unstable states play the role of way-stations, near which chaotic or turbulent trajectories spend much of their time. These are believed to form the core structures supporting weakly turbulent dynamics. Among the unstable periodic orbits that may influence trajectories of a fluid-dynamical system, we have discovered some whose branches terminate in global bifurcations, leading to their disappearance. Although there have been a number of computations of global bifurcations in hydrodynamic systems \citep{Tuckerman1988global, prat2002, Nore2003mode, millour2003sensitivity, abshagen2005symmetry, bordja2010influence, bengana2019spirals, Reetz2020b}, we are not aware of previous calculations of heteroclinic or homoclinic cycles in vertical convection.

\par The remainder of this manuscript is organised as follows: in \S 2 we summarize the key numerical methods used in our research which are already presented in detail in \citet{Zheng2023part1}. The results from the bifurcation analysis will be shown in \S 3 for fixed points and in \S 4 for periodic orbits. Concluding remarks and future research directions will be outlined in \S 5.

\section{System and methods}
\par We refer readers to \citet{ReetzPhD, Reetz2020a, Reetz2020b, Zheng2023part1} for detailed descriptions of the numerical methods used in the research. Here, we will only summarize the key points.

\begin{figure}
    \centering
    \includegraphics[width=0.5\columnwidth]{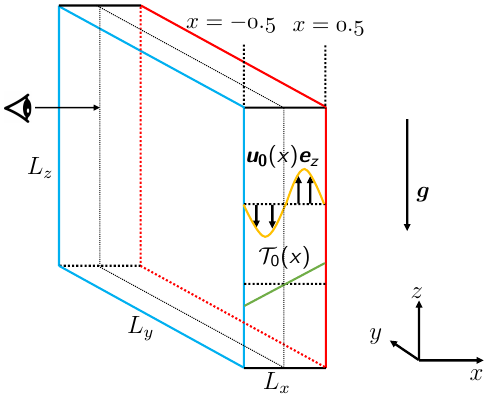}
    \captionsetup{font={footnotesize}}
    \captionsetup{width=13.5cm}
    \captionsetup{format=plain, justification=justified}
    \caption{\label{ILC_figure} Schematic of the vertical convection configuration approximating $[L_x, L_y, L_z] = [1, 8, 9]$. The flow is bounded between two walls in $x$ direction at $x=0.5$ where the flow is heated and at $x=-0.5$ where the flow is cooled. The domain is periodic in $y$ and $z$ directions. Most of the visualizations that we present are taken on the $y$-$z$ midplane at $x=0$ outlined by the dotted lines, and they are visualized in the direction of negative to positive $x$, as indicated by the eye and arrow. The laminar velocity and temperature are shown as the orange and green curves, respectively.}
\end{figure}

\subsection{Direct numerical simulation of vertical convection}
\par The vertical convection system is studied numerically by performing direct numerical simulations (DNS) with the ILC extension module of the \textit{Channelflow 2.0} code \citep{Gibson2019}, to solve the non-dimensionalized Oberbeck–Boussinesq equations:
\begin{subequations}
\label{appabc}
\begin{align}
    \dfrac{\partial \boldsymbol u}{\partial t} + (\boldsymbol u \cdot \nabla)\boldsymbol u &= -\nabla p + \left(\frac{Pr}{Ra}\right)^{1/2} \nabla^2 \boldsymbol u + \mathcal{T}\boldsymbol e_z, \label{appa} \\
    \dfrac{\partial \mathcal{T}}{\partial t} + (\boldsymbol u \cdot \nabla)\mathcal{T} &= \left(\frac{1}{Pr\: Ra}\right)^{1/2}\nabla^2 \mathcal{T}, \label{appb} \\
    \nabla \cdot \boldsymbol u&= 0,	\label{appc} 
\end{align}
\end{subequations}
in a vertical channel, with periodic boundary conditions in $y$ and $z$, shown in figure \ref{ILC_figure}. The boundary conditions in $x$ at the two walls are of Dirichlet type:
\begin{align}
    \boldsymbol u(x=\pm 0.5) = 0,
    \qquad 
    \mathcal{T}(x=\pm0.5)=\pm0.5. 
\label{eq:bcs}
\end{align} 
Supplementary integral constraints are necessary in the periodic directions; we set the mean pressure gradient to zero in both $y$ and $z$. The laminar solution, illustrated in figure \ref{ILC_figure}, is:
\begin{subequations}
\label{laminarabc}
\begin{align}
    &\boldsymbol u_0(x) = \frac{1}{6}\sqrt{\frac{Ra}{Pr}} \left(\frac{1}{4}x - x^3\right) \boldsymbol e_z,	\label{laminara} \\
    &\mathcal{T}_0(x) = x,  \label{laminarb} \\
    &p_0(x) = \Pi,	\label{laminarc} 
\end{align}
\end{subequations}
with arbitrary pressure constant $\Pi$. 

\par The governing equations and boundary conditions are discussed in our companion paper \cite{Zheng2023part1}. The only aspect which differs here is the domain size: instead of the narrow domain $[L_x, L_y, L_z] = [1, 1, 10]$ with one extended direction studied in \citet{Gao2013}, here we study the three-dimensional computational domain $[L_x, L_y, L_z] = [1, 8, 9]$ of \citet{Gao2018}. This domain has two extended directions and is illustrated in figure \ref{ILC_figure}. This domain is spatially discretized by $[N_x, N_y, N_z] = [31, 96, 96]$ Chebychev-Fourier–Fourier modes.

\subsection{Symmetries and computation of invariant solutions}
\par We will often refer to the symmetries of our system, the group $S_{VC}$, which is generated by reflection in $y$, combined reflection of $x$, $z$ and temperature $\mathcal{T}$, and translation in $y$ and $z$:
\begin{subequations}
    \begin{eqnarray}
        &\pi_y[u,v,w,\mathcal{T}](x,y,z) \equiv [u,-v,w,\mathcal{T}](x, -y,z) \label{sym_a},\\
        &\pi_{xz}[u,v,w,\mathcal{T}](x,y,z) \equiv [-u,v,-w,-\mathcal{T}](-x, y,-z) \label{sym_b}, \\
        &\tau(\Delta y, \Delta z)[u,v,w,\mathcal{T}](x,y,z) \equiv [u,v,w,\mathcal{T}](x, y + \Delta y,z + \Delta z), \label{sym_c}
    \end{eqnarray}
\end{subequations}
stated more compactly as $S_{VC} \equiv \braket{\pi_y, \pi_{xz}, \tau(\Delta y, \Delta z)} \simeq [O(2)]_y \times [O(2)]_{x,z}$. The groups we use are $Z_n$, the cyclic group of $n$ elements, $D_n$, the cyclic group of $n$ elements together with a non-commuting reflection, and $O(2)$, the group of all rotations together with a non-commuting reflection. $[O(2)]_y$ refers to reflections and translations in $y$, as in \ref{sym_a} and \ref{sym_c}, respectively, while $[O(2)]_{xz}$ refers to reflections in $(\mathcal{T},x,z)$ as in \ref{sym_b} and translations in $z$ as in \ref{sym_c}, a convention that we will use in the rest of the paper where possible. Note that the generators of a group are non-unique, as is the decomposition into direct products (indicated by $\times$). 

\par We adopt the shooting-based matrix-free Newton method implemented in \textit{Channelflow 2.0} to compute invariant solutions. The only difference with respect to our description in \cite{Zheng2023part1} arises from the presence here of homoclinic and heteroclinic orbits. While the Newton method can converge with one shot in most of the cases (provided that the initial guess is sufficiently close to the solution), the multi-shooting method \citep{Veen2010, Sanchez2010} is required in order to converge orbits with long periods (typically $T > 300$ in our case) that are close to a global bifurcation point and very unstable orbits. For these periodic orbits, we employ the multi-shooting method with at most six shots. 

\par To characterise the stability of a solution, its leading eigenvalues and eigenvectors for fixed points, or Floquet exponents and Floquet modes for periodic orbits, are determined by Arnoldi iterations. When solutions have symmetries, the resulting linear stability problem has the same symmetries, leading to multiple eigenvectors sharing the same eigenvalues. In such cases, we choose the eigenvectors appropriate to our analysis either by subtracting two nonlinear flow fields along a trajectory or branch, or by imposing symmetries.

\subsection{Order parameter and flow visualization}
\par Once an equilibrium or time-periodic solution is converged, parametric continuation in Rayleigh number is performed to construct bifurcation diagrams. Solutions are represented via the $L_2$-norm of their temperature deviation $\theta \equiv \mathcal{T} - \mathcal{T}_0$. Branches of fixed points are represented by curves showing $\lvert\lvert \theta \lvert\lvert_2$ as a function of $Ra$; for periodic orbits, the maximum and minimum of $\lvert\lvert \theta \lvert\lvert_2$ along an orbit are plotted. The thermal energy input $I$ due to buoyancy and the dissipation $D$ due to viscosity are used to plot phase portraits.

\section{Fixed points}
\label{part2_Equilibria}
\par We begin by noting that the numbering used for fixed points and for periodic orbits applies only to this paper; except for FP1, the fixed points and periodic orbits here are not the same as those in \citet{Zheng2023part1}. 

\begin{figure}
    \centering
    \includegraphics[width=\columnwidth]{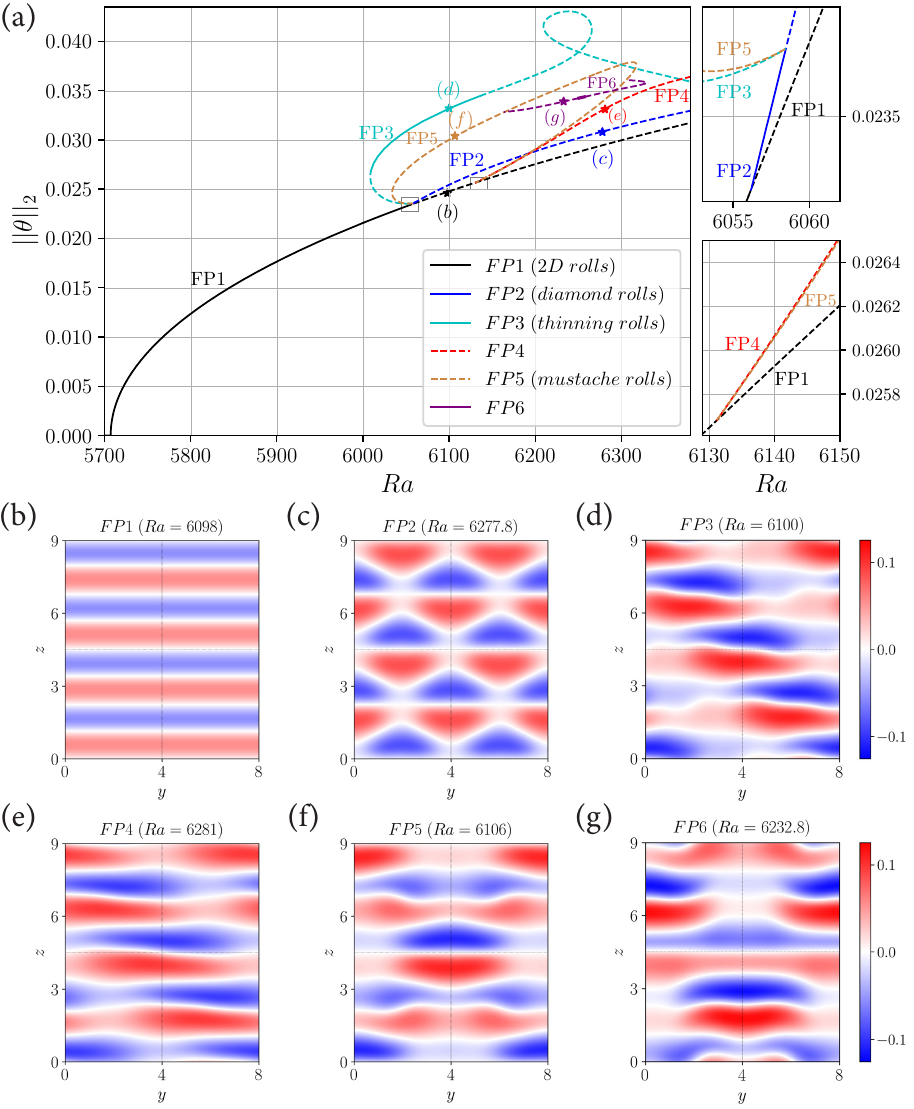}
    \captionsetup{font={footnotesize}}
    \captionsetup{width=13.5cm}
    \captionsetup{format=plain, justification=justified}
    \caption{\label{FP918} Bifurcation diagram (a) and flow structures visualized via the temperature field on the $y$-$z$ plane at $x=0$ (b-g) of six equilibria in domain $[L_x, L_y, L_z] = [1, 8, 9]$. The diagram shows two supercritical pitchfork bifurcations, one from the base state to FP1 (b) and another one from FP1 to FP2 (c). FP3 (d) bifurcates from FP2 in a subcritical pitchfork bifurcation. The unstable FP4 (e) bifurcates supercritically from FP1. The unstable FP5 branch (f) bifurcates at one end subcritically from FP2, and at the another end supercritically from FP1. FP3 and FP5 bifurcate together from FP2, while FP4 and FP5 bifurcate together from FP1. The two grey rectangles surround these two simultaneous bifurcations, which are also shown in the enlarged diagrams on the right. On the lower zoomed diagram, the dashed red and brown lines are distinct, but too close to one another to be distinguished. FP6 bifurcates from FP5 in two supercritical pitchfork bifurcations and it connects FP5 at two Rayleigh numbers. In (a), solid and dashed curves signify stable and unstable states respectively. The ranges over which FP1, FP2, FP3 and FP6 are stable are $[5707, 6056]$, $[6056, 6058.5]$, $[6008.5, 6140]$ and $[6251.4, 6257.6]$, respectively. The stars in (a) indicate the locations of the visualisations of (b-g). FP1-FP3 are discussed in \citet{Gao2018} while FP4-FP6 are newly identified in this work. Other branches of equilibria exist, which we have not followed nor shown on this diagram. Flow visualizations on the $x$-$z$ plane are shown in figure \ref{FP918_xz}.}
\end{figure}
\begin{figure}
    \centering
    \includegraphics[width=\columnwidth]{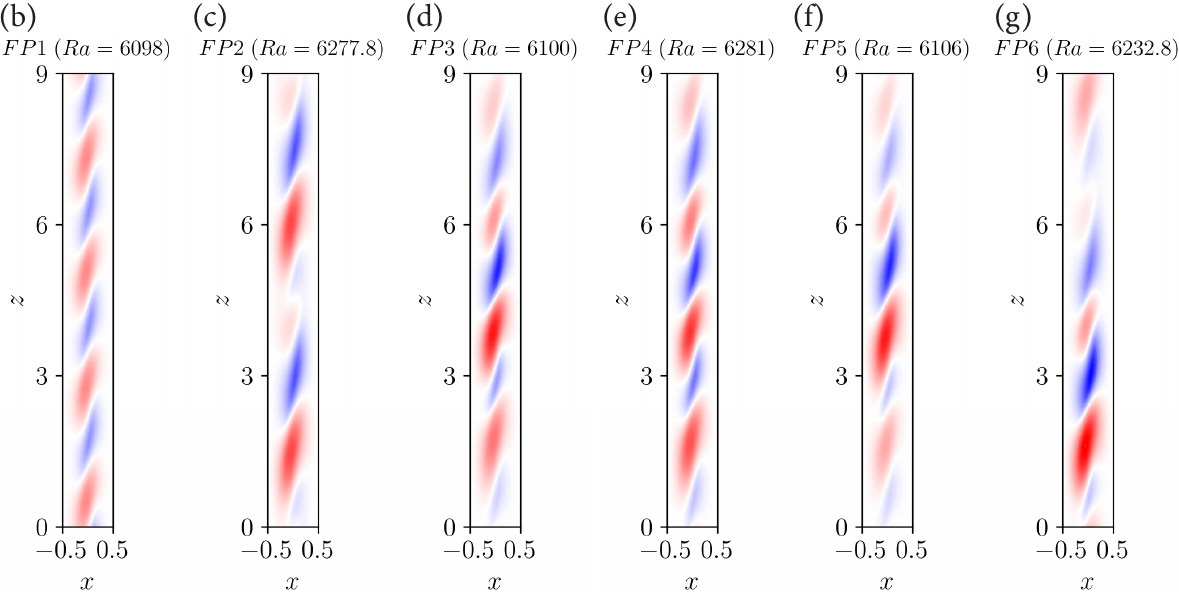}
    \captionsetup{font={footnotesize}}
    \captionsetup{width=13.5cm}
    \captionsetup{format=plain, justification=justified}
    \caption{\label{FP918_xz} Flow visualization complementary to figure \ref{FP918}(b-g): FP1-FP6 visualized via the temperature field on the $x$-$z$ plane at $y=4$. The same color bar is used as in figures \ref{FP918}(b-g).}
\end{figure}

\subsection{Three known fixed points: FP1-FP3}
\label{part2_known_FP}
\par \citet{Gao2018} observed three fixed points in the domain $[L_x, L_y, L_z] = [1, 8, 9]$ and presented visualisations and Fourier decompositions of them. These states have been recomputed here and their flow structures are shown in figures \ref{FP918}(b-d). In this work, we identify the bifurcations that create and destroy these states and construct a bifurcation diagram that includes stable and unstable branches. As presented in the bifurcation diagram in figure \ref{FP918}(a), the laminar base flow is stable until $Ra=5707$, where the first fixed point, FP1, bifurcates. As in \citet{Zheng2023part1}, FP1 is called two-dimensional or transverse rolls. This state contains four spanwise ($y$)-independent co-rotating convection rolls, and is shown in figures \ref{FP918}(b) and \ref{FP918_xz}(b). \cite{cingi2021direct} have reported bistability between the base flow and two-dimensional rolls in several Rayleigh-number ranges, but their interpretation contradicts the results obtained here and also those reported by \cite{Gao2018}. In particular, \cite{cingi2021direct} find the laminar flow to be bistable with 2D rolls (FP1) over the Rayleigh number range of $[5708, 7000]$. We believe this reported bistability to be spurious, and to almost certainly result from the use by \cite{cingi2021direct} of a time-stepping code to simulate a weakly unstable state without monitoring the growth or decay of perturbations nor a complementary linear stability analysis.

\par FP1 loses stability at $Ra = 6056$ via a circle pitchfork bifurcation that breaks the $y$ translation symmetry $\tau(\Delta y,0)$ and creates FP2, shown in figures \ref{FP918}(c) and \ref{FP918_xz}(c). We will refer to these as diamond rolls, while \cite{Gao2018} called them wavy rolls. FP2 results from the subharmonic varicose instability of FP1 and this instability is discussed in \citet{Subramanian2016, Reetz2020b}. FP2 undergoes subcritical pitchfork bifurcations at $Ra = 6058.5$, so that its stability range is only $[6056,6058.5]$. The time-dependent simulations of \cite{cingi2021direct} did not detect FP2. In contrast, \cite{Gao2018} observed FP2 as a transient at $Ra=6100$ and computed its threshold via a linear stability analysis. \cite{Clever1995} computed a state resembling FP2 by means of a steady-state calculation. (Their threshold of about $Ra\approx6295$ can perhaps be attributed to a lack of spatial resolution available in 1995.)

\par The bifurcation from FP2 creates FP3, which \cite{Gao2018} call thinning rolls. Initially unstable, FP3 is stabilized by a saddle-node bifurcation at $Ra = 6008.5$. At higher Rayleigh number, FP3 undergoes two additional saddle-node bifurcations at $Ra=6265.8$ and $Ra=6209.56$. As pointed out by \citet{Gao2018}, FP3 can have either of two possible diagonal orientations. Figure \ref{FP918}(d) shows one of the two cases: the slightly wider red portions are located along a diagonal joining the top left with the bottom right. 

\par The symmetry (isotropy) groups of FP1-FP3 are
\begin{equation}
\begin{array}{lll}
\text{FP1:} \; & \braket{\pi_y, \tau(\Delta y,0),\pi_{xz},\tau(0,L_z/4)} & \simeq [O(2)]_y \times [D_4]_{xz};\\
\text{FP2:} \; & \braket{\pi_y, \tau(L_y/2,0), \pi_{xz}, \tau(0, L_z/2), \tau(L_y/4, -L_z/4)} & \simeq D_2 \times D_4; \\ 
\text{FP3:} \; & \braket{\pi_y\pi_{xz}, \tau(L_y/4,-L_z/4)} & \simeq D_4.
\end{array}
\end{equation}
Note that $\tau(L_y/4,-L_z/4)=\tau(L_y/4,3L_z/4)$, and that the symmetry groups for FP2 and FP3 cannot be divided into those related to $y$ and those related to $x,z$. The bifurcation from FP2$\rightarrow$FP3 breaks the $D_4$ symmetry of FP2.

\subsection{Three new fixed points: FP4-FP6}
\label{largedomain_newFP}
\par We have also found three new branches of fixed points, FP4-FP6. Figure \ref{FP918}(a) shows that there is a supercritical pitchfork bifurcation at $Ra=6131$, at which FP4 and FP5 bifurcate simultaneously from FP1. Both FP4 and FP5 are unstable along their entire branches. (The enlarged diagram on the bottom right of figure \ref{FP918}(a) contains two distinct dashed red and brown lines which are too close to be distinguished.) Since FP1 is $y$-independent and FP4 and FP5 are not, these are circle pitchfork bifurcations, yielding FP4 and FP5 states of any phase in $y$. FP4, shown in figures \ref{FP918}(e) and \ref{FP918_xz}(e), shares with FP3 a diagonal orientation. FP4 also consists of rolls with a slight wavy modulation along the $y$ direction, but this modulation is weaker than that of FP3. FP4 plays an essential role in one of the global bifurcations that we will discuss in \S \ref{homo_termination}. 

\par The FP5 branch (which we will refer to occasionally as the mustache branch) is shown in figures \ref{FP918}(f) and \ref{FP918_xz}(f). After bifurcating from FP1, the FP5 branch undergoes saddle-node bifurcations at $Ra=6317.5$ and $Ra= 6034$, towards decreasing and increasing Rayleigh number, respectively, and finally terminates at $Ra=6058.5$ by meeting FP2 in a subcritical pitchfork bifurcation. This is not a circle pitchfork bifurcation, since the diamond branch FP2 is also $y$-dependent; four possible FP5 branches emanate from FP2, related to one another by translations in $y$ and in $z$. (FP3 is also created at $Ra=6058.5$, in another simultaneous bifurcation that will be discussed in \S \ref{two_simu_bif_part2}.) Thus, two routes connect FP1 to FP5: a single circle pitchfork bifurcation, and a circle pitchfork bifurcation from FP1 to FP2 followed by an ordinary pitchfork bifurcation from FP2 to FP5. The bifurcation from FP1 to FP2 breaks $y$ invariance while that from FP2 to FP5 breaks the four-fold translation symmetry $\tau(L_y/4,-L_z/4)$.

\par The last new equilibrium, FP6, shown in figures \ref{FP918}(g) and \ref{FP918_xz}(g), is created from FP5 at $Ra=6164.3$ in a supercritical pitchfork bifurcation, inheriting the instability of FP5 at the bifurcation point. FP6 becomes stable, but only over a very short range $Ra \in [6251.4,6257.6]$, indicated by the slight thickening of the branch in figure \ref{FP918}(a). (We do not discuss or show in figure \ref{FP918}(a) the new branches that necessarily emanate from the stabilising bifurcation at $Ra=6251.4$, nor the numerous other branches created at points at which the real part of an eigenvalue crosses zero. The bifurcation at $Ra=6257.6$ will be discussed in \S \ref{PO4_Hopf}.) FP6 then undergoes a saddle-node bifurcation at $Ra=6329$ before terminating at the FP5 branch at $Ra=6305.8$ in another supercritical pitchfork bifurcation. 

\par The symmetry groups of these states are 
\begin{equation}
\begin{array}{lll}
\text{FP4:} & \braket{\pi_y\pi_{xz}, \tau(L_y/4,-L_z/4)} &\simeq D_4; \\
\text{FP5:} & \braket{\pi_y,\pi_{xz}, \tau(L_y/2, L_z/2)} & \simeq [Z_2]_y \times [Z_2]_{xz} \times Z_2; \\ 
\text{FP6:} & \braket{\pi_y,\pi_{xz}\tau(L_y/2,0)} &\simeq [Z_2]_y\times Z_2.
\end{array}
\end{equation}

\par FP1 is homogeneous in $y$ and the states which branch from it, directly or indirectly, are FP2 with a $y$ periodicity of $L_y/2=4$, and FP3, FP4, FP5, and FP6 with $y$-periodicity $L_y=8$. This sets the stage for 1:2 mode interaction, as analysed in detail by \citet{Armbruster1988}, one of whose consequences is a robust heteroclinic cycle to be discussed in \S \ref{robust_hetero_cycle}.

\subsection{Two simultaneous bifurcations}
\label{two_simu_bif_part2}
\begin{figure}
    \centering
    \includegraphics[width=\columnwidth]{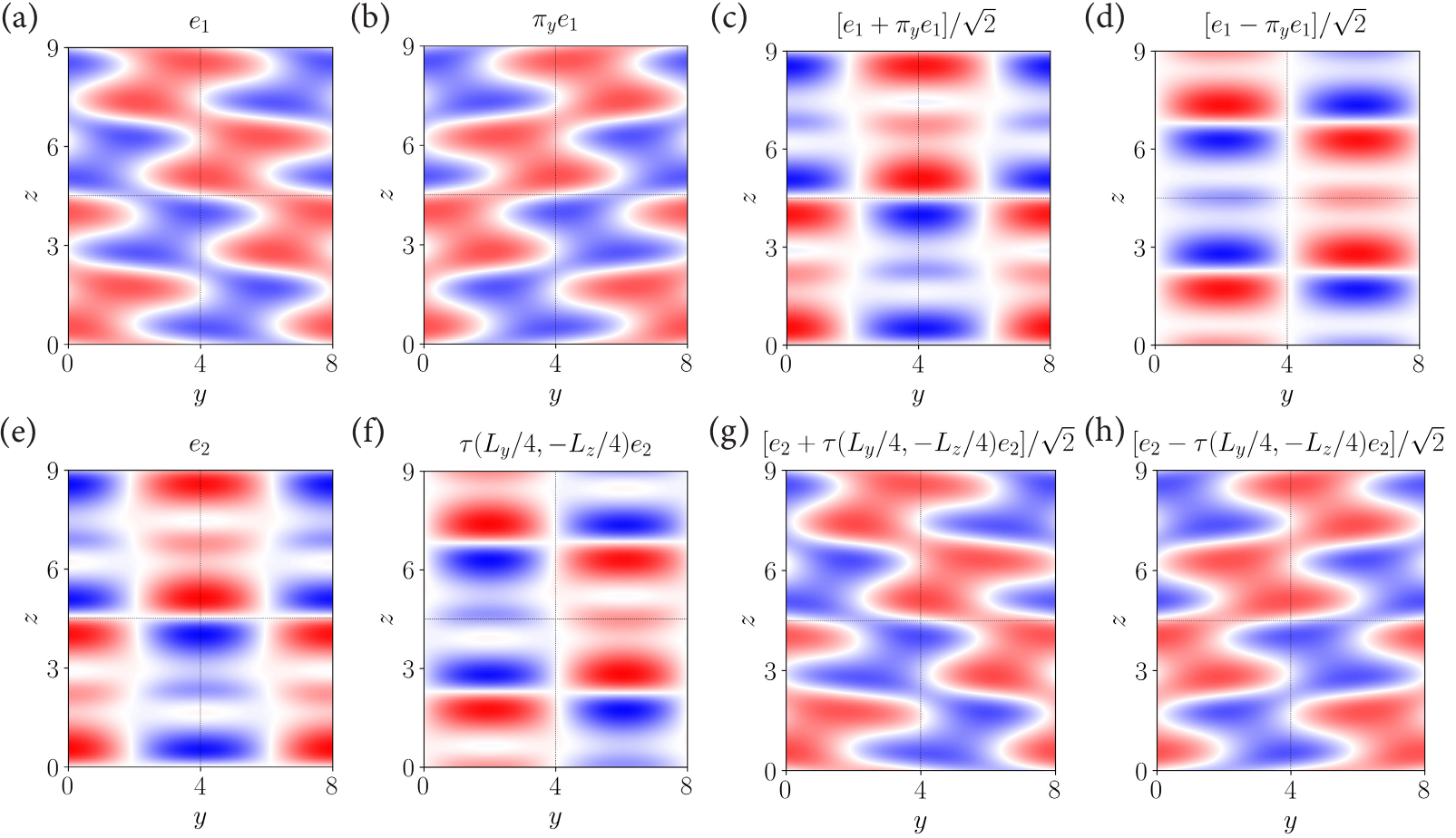}
    \captionsetup{font={footnotesize}}
    \captionsetup{width=13.5cm}
    \captionsetup{format=plain, justification=justified}
    \caption{\label{D4_FP2} (a) Eigenvector $e_1$ responsible for FP2$\rightarrow$FP3 bifurcation (obtained by subtracting FP2 at $Ra=6058.5$ from FP3 at $Ra=6056$) and (b) its $y$-reflected version $\pi_y e_1$. (c-d) Superpositions $(e_1 \pm \pi_y e_1)/\sqrt{2}$. (e) Eigenvector $e_2$ responsible for FP2$\rightarrow$FP5 bifurcation (obtained by subtracting FP2 at $Ra=6058.5$ from FP5 at $Ra=6056$) and (f) its quarter-diagonal translation $\tau(L_y/4,-L_z/4) e_2$. (g-h) Superpositions $(e_2 \pm \tau(L_y/4,-L_z/4) e_2)/\sqrt{2}$. All eigenvectors are visualized via the temperature field on the $y$-$z$ plane at $x=0$. The same color bar is used in all plots.}
\end{figure}

\par The two enlarged bifurcation diagrams on the right of figure \ref{FP918}(a) depict bifurcations at which two qualitatively different branches with different symmetries are created simultaneously. FP3 and FP5 bifurcate simultaneously from FP2 at $Ra=6058.5$, and FP4 and FP5 bifurcate simultaneously from FP1 at $Ra= 6131$. These simultaneous bifurcations can be explained by the same $D_4$ scenario that is discussed in detail in \citet{Zheng2023part1}. We repeat here the normal form corresponding to bifurcation in the presence of $D_4$ symmetry:
\begin{subequations}
\label{D4_eqns}
\begin{align}
\dot{p} &= \left(\mu - a p^2 - b q^2\right) p,\\
\dot{q} &= \left(\mu - b p^2 - a q^2\right) q.
\end{align}
\end{subequations}
The dynamical system \eqref{D4_eqns} has the non-trivial solutions
\begin{subequations}
\label{D4_sol}
\begin{align}
  p &= \pm\sqrt{\mu/a} & q &= 0, \label{D4_sola}\\
  p &= 0 & q&=\pm\sqrt{\mu/a}, \label{D4_solb}\\
  p &= \pm\sqrt{\mu/(a+b)} & q& = \pm\sqrt{\mu/(a+b)}, \label{D4_solc}\\
  p &= \pm\sqrt{\mu/(a+b)} & q& = \mp\sqrt{\mu/(a+b)}, \label{D4_sold}
\end{align}
\end{subequations}
i.e.\ two classes of solutions, \eqref{D4_sola}-\eqref{D4_solb}, which we call here the diagonal solutions, and \eqref{D4_solc}-\eqref{D4_sold}, which we call here the rectangular solutions, for reasons which figure \ref{D4_FP2} will make clear. The diagonal solutions are related to one another by symmetry, as are the rectangular ones, but the diagonal solutions are not related by symmetry to the rectangular solutions. 

\par We begin by explaining the simultaneous bifurcation from FP2. The symmetry group $D_4$ of FP2 is generated by the translation operator $\tau(L_y/4, -L_z/4)$ together with either of the reflection operators, $\pi_y$ or $\pi_{xz}$. FP2 is invariant under any product of these operations. In the model \eqref{D4_eqns}, FP2 corresponds to the trivial solution $p=q=0$ from which the other solutions bifurcate. 

\par When FP2 loses stability at $Ra=6058.5$, a real eigenvalue $\lambda_{1,2}$ crosses the imaginary axis. This double eigenvalue has a two-dimensional eigenspace, spanned by any two of its linearly independent eigenvectors. Figure \ref{D4_FP2}(a) shows the eigenvector $e_1$ of FP2 giving rise to state FP3 shown in figure \ref{FP918}(d), while figure \ref{D4_FP2}(b) shows its $y$-reflection, $\pi_y e_1$. Since $\pi_y$ belongs to the symmetry group of FP2, $\pi_y e_1$ is also an eigenvector of FP2, as is any superposition of $e_1$ and $\pi_y e_1$. The diagonal solution \eqref{D4_sola} represents FP3 which arises from eigenvector $e_1$. Solution \eqref{D4_solb} represents FP3$^\prime \equiv \pi_y$FP3, whose diagonal is reversed and which arises from eigenvector $\pi_y e_1$. The amplitudes of $e_1$ and $\pi_y e_1$ are represented in the model \eqref{D4_eqns} by variables $p$ and $q$:
\begin{align}
{\rm FP3} = {\rm FP2} + p(t) e_1 + q(t) \pi_y e_1.
\end{align}

\par The eigenvector $e_2$ of FP2 leading to state FP5 is shown in figure \ref{D4_FP2}(e). Eigenvector $e_2$ turns out to be identical to the equal superposition of $e_1$ and $\pi_y e_1$, as shown in figure \ref{D4_FP2}(c). This is a manifestation of the fact that, in the model \eqref{D4_eqns}, the rectangular solutions \eqref{D4_solc} and \eqref{D4_sold} contain equal amplitudes of $p$ and $q$. The shifted eigenvector $\tau(L_y/4,-L_z/4)e_2$ leads to FP5$^{\prime} \equiv \tau(L_y/4,-L_z/4)\:$FP5; its superposition with $e_2$ produces $e_1$. Indeed, in the model \eqref{D4_eqns}, equal superpositions of rectangular solutions of types \eqref{D4_solc} and \eqref{D4_sold} produce the diagonal solutions of types \eqref{D4_sola} and \eqref{D4_solb}. (Figures \ref{D4_FP2}(d) and (h) are also eigenvectors of FP2, identical to figures \ref{D4_FP2}(f) and (b), respectively.) In figure \ref{D4_FP2}, the eigenvectors have been approximated by subtracting FP2 from FP3 and from FP5 just beyond the bifurcation point ($Ra=6058.5$ for FP2 and $Ra=6056$ for FP3 and FP5). This selects the appropriate choices out of the multitude of eigenvectors of the highly symmetric FP2. 

\par Just as solutions \eqref{D4_solc} and \eqref{D4_sold} are not related to solutions \eqref{D4_sola} or \eqref{D4_solb} by any symmetry operation, FP5 cannot be produced by a symmetry transformation from FP3. In addition, figure \ref{FP918}(a) makes it clear that branches FP3 and FP5 behave differently, with a different global temperature norm and different saddle-node bifurcations. 

\par We turn now to the simultaneous bifurcations of FP4 and FP5 from FP1 at $Ra = 6131$. The symmetries of FP1 are generated by reflection and translation in $y$ together with reflection in $(x,z)$ and four-fold-translation in $z$, i.e. $[O(2)]_y \times [D_4]_{xz}$. We again compute the eigenvectors of FP1 responsible for these two bifurcations. Taking symmetry transformations and superpositions, we obtain the eigenvector responsible for FP5 (FP4) as the equal superposition of the eigenvector responsible for FP4 (FP5) with a symmetry-transformed version of it. Interestingly, the eigenvectors responsible for the simultaneous bifurcation from FP1$\rightarrow$(FP4, FP5) at $Ra=6131$ are very similar to those responsible for the simultaneous bifurcation from FP2$\rightarrow$(FP3, FP5) at $Ra=6058.5$. This can be explained as follows. The two simultaneous bifurcations occur at Rayleigh numbers which are close to each other and to $Ra=6056$, at which FP2 is formed via a supercritical circle pitchfork bifurcation from FP1. FP2 inherits the spectrum of FP1, with the exception of the double eigenvalue responsible for the circle pitchfork. (Just above $Ra=6056$, this double eigenvalue becomes positive for FP1, whereas it splits into a zero and negative eigenvalue for FP2.) The other eigenvectors and eigenvalues of FP2 at $Ra=6058.5$ are close to those of FP1 at $Ra=6131$, including those shown in figure \ref{D4_FP2} which cause the simultaneous bifurcations. We do not show the eigenvectors of FP1 to avoid repetition.

\par It has been known since the mid-1980s \citep{swift1985bifurcation} that $D_4$ symmetry leads to the simultaneous creation of non-symmetry-related branches. This has been applied to a number of situations, such as the simultaneous creation of standing and traveling waves \citep{Knobloch1986, Boronska2006, Reetz2020b}. The application most relevant here is that of counter-rotating Taylor-Couette flow, in which spirals were first described in the classic paper by \citet{Taylor1923}. The superposition of spirals of opposite helicity leads to a state called ribbons, much as the superposition of diagonal states produces the rectangular states in the current study. Exceptionally, ribbons were first predicted mathematically \citep{demay1984calcul, Chossat1994}, setting off a quest to observe them experimentally, which was finally achieved by \citet{tagg1989nonlinear}.

\section{Periodic orbits}
\label{part2_PO}
\begin{figure}
    \centering
    \includegraphics[width=\columnwidth]{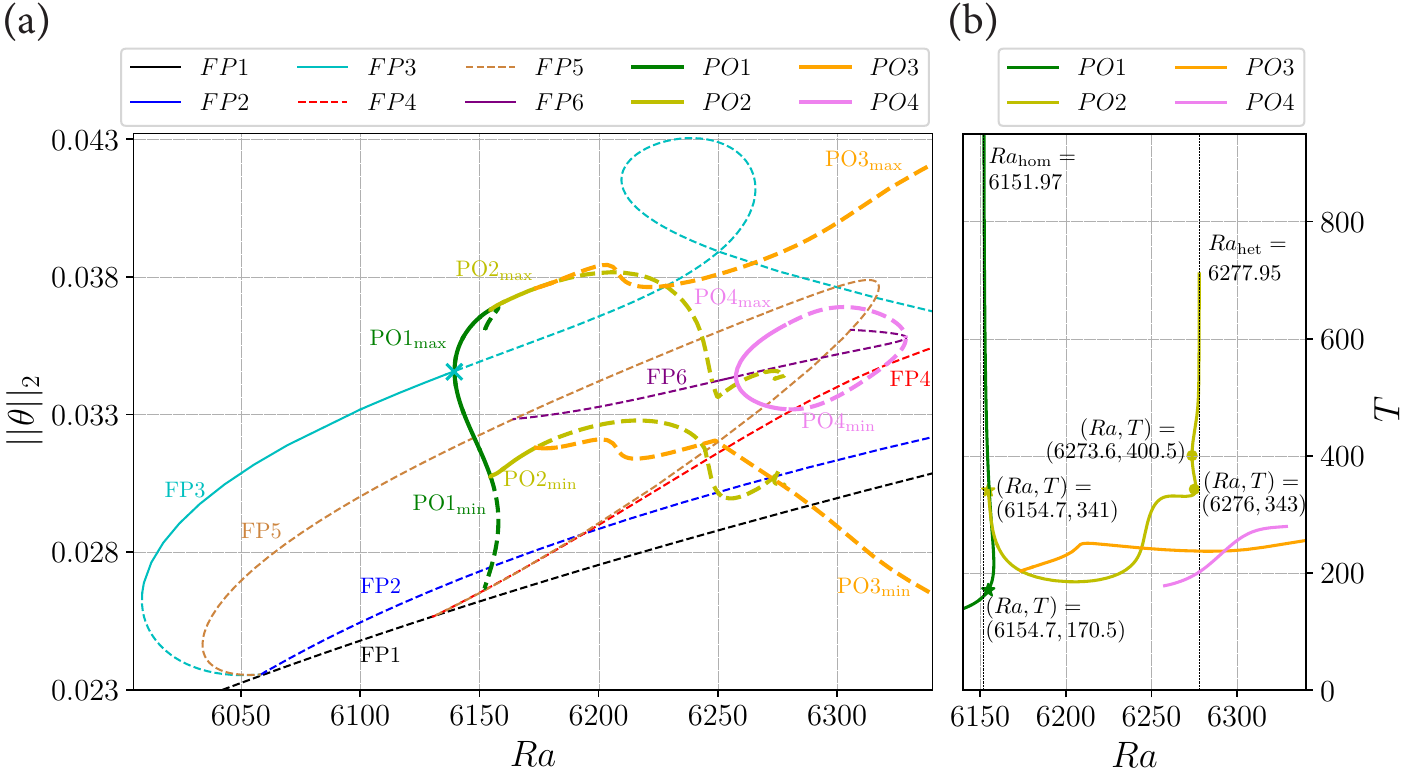}
    \captionsetup{font={footnotesize}}
    \captionsetup{width=13.5cm}
    \captionsetup{format=plain, justification=justified}
    \caption{\label{largedomain_BDall} (a) Bifurcation diagram of fixed points (FPs) and periodic orbits (POs) and (b) periods of four periodic orbits in domain $[L_x, L_y, L_z] = [1, 8, 9]$. In (a), for each periodic orbit, we show two curves, the maximum and minimum of $\lvert\lvert \theta \lvert\lvert_2$ along an orbit. PO1 appears via a Hopf bifurcation from FP3 at $Ra=6140$ (marked by a cyan cross) and undergoes a period-doubling bifurcation at $Ra=6154.7$ giving rise to PO2. PO1 then undergoes a saddle-node bifurcation at $Ra=6157.97$ and disappears by meeting FP4 in a homoclinic bifurcation at $Ra_{\rm hom}=6151.97$ at which its period diverges; see (b). PO2 loses stability at $Ra = 6173.8$ where PO3 is created via a supercritical pitchfork bifurcation. The stability of PO2 changes multiple times along the branch for $6235<Ra<6255$, see details in figure \ref{evo_eigen_PO2}. PO2 then undergoes two closely spaced saddle-node bifurcations (at $Ra=6276$ and $6273.6$; see (b)) before terminating by meeting two symmetrically-related versions of FP2 in a heteroclinic bifurcation at $Ra_{\rm het}=6277.95$, at which its period diverges. PO3 is continued until $Ra=6407.3$ (the range $6340<Ra<6407.3$ is not shown) and its period remains approximately constant. The apparent lack of smoothness in the curves representing PO2 and PO3 (in (a) around $Ra=6250$) corresponds to the overtaking of one temporal maximum or minimum of $\lvert\lvert \theta \lvert\lvert_2$ by another as $Ra$ is varied. PO4 bifurcates from and terminates on FP6 at $Ra=6257.6$ and $Ra=6328.8$, and it is stable within $6257.6<Ra<6278$. In (a), solid and dashed curves signify stable and unstable states respectively, and the curves representing periodic orbits are slightly thicker than those of fixed points. The same color code is used in (a) and (b). A schematic bifurcation diagram is shown in figure \ref{Schematic-BD}. Many other branches of equilibria and periodic orbits exist, which we have not followed nor shown on this diagram.}
\end{figure}

\par In this section, we explore four periodic orbits, PO1 to PO4. Periodic orbits PO1 to PO3 are created by a sequence of local bifurcations (i.e.\ bifurcations associated with a change in the real part of one or more eigenvalues/Floquet exponents): FP3$\rightarrow$PO1$\rightarrow$PO2$\rightarrow$PO3. PO1 and PO2 disappear in a global homoclinic and heteroclinic bifurcation, respectively, while the termination of PO3 is not discussed in this work. PO4 bifurcates from and terminates on FP6 via Hopf bifurcations. The bifurcation diagram of figure \ref{largedomain_BDall}(a) shows the six equilibria discussed in \S \ref{part2_Equilibria} and the four periodic orbits to be discussed, while the periods of the limit cycles are shown in figure \ref{largedomain_BDall}(b). 

\subsection{First periodic orbit (PO1)}
\subsubsection{Creation of PO1: Hopf bifurcation}
\begin{figure}
    \centering
    \includegraphics[width=\columnwidth]{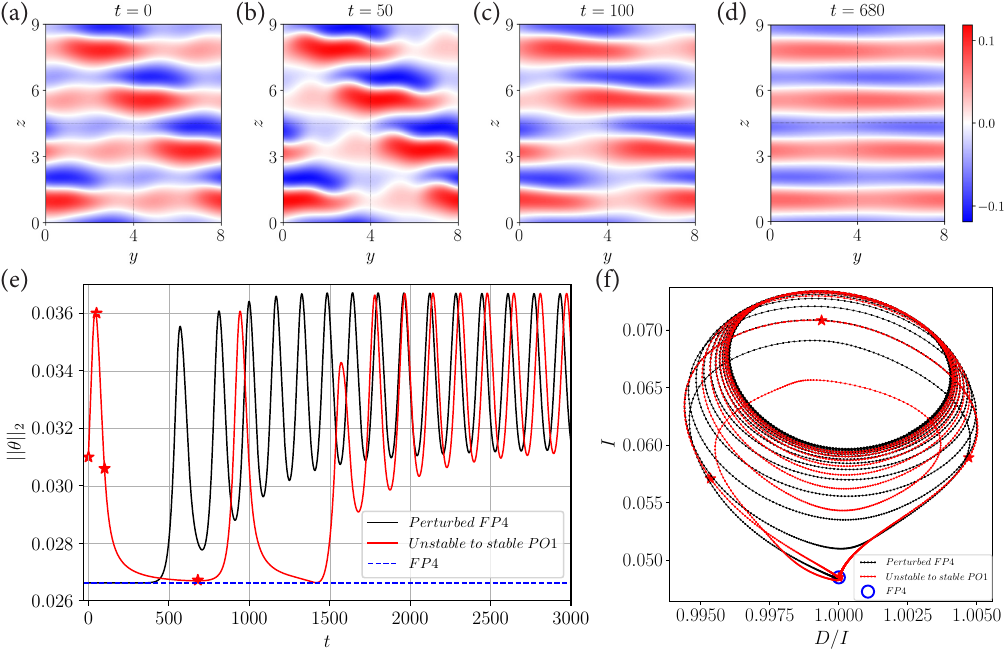}
    \captionsetup{font={footnotesize}}
    \captionsetup{width=13.5cm}
    \captionsetup{format=plain, justification=justified}
    \caption{\label{global_bifurcation} (a-d) The dynamics of PO1 (visualized via the temperature field on the $y$-$z$ plane at $x=0$) on the unstable lower branch at $Ra = 6152.249$ ($Ra_{\rm hom} = 6151.97$). Snapshot (d) converges to FP4 when used as an initial estimate for Newton solving. (e) Time series from DNS at $Ra =6152.249$ ($T=900$), initialized by the unstable PO1 shown in (a) (red curve) and by FP4 with a small perturbation (black). The trajectory initialized by the unstable PO1 spends a long time near FP4 ($250<t<800$). Both simulations converge to the stable PO1 branch ($t>2500$) at this Rayleigh number. (f) Phase portrait illustrating the same data set as in (e). The plot shows the thermal energy input ($I$) versus the viscous dissipation over energy input ($D/I$). FP4 (hollow blue circle) is located on the vertical line $D/I=1$, where energy dissipation and input are equal. The four red stars in (e) and (f) indicate the moments at which the snapshots (a)-(d) are taken. The same color code is used in (e) and (f).}
\end{figure}

\par Produced by a subcritical pitchfork bifurcation from FP2, FP3 is unstable at onset, but is stabilized by a saddle-node bifurcation at $Ra = 6008.5$ and then loses stability again at $Ra = 6140$ via a supercritical Hopf bifurcation that produces a periodic orbit PO1. PO1 inherits all of the spatial symmetries of FP3: $\braket{\pi_y\pi_{xz}, \tau(L_y/4,-L_z/4)} \simeq D_4$, and hence no additional spatio-temporal symmetries are present. The S-shaped green curve in figure \ref{largedomain_BDall}(a) contains the maximum (PO1$_{max}$, above the cyan curve of FP3) and minimum (PO1$_{min}$, below the cyan curve) values of $\lvert\lvert \theta \lvert\lvert_2$ over the period of each PO1 state. The period ($T$) of PO1 increases smoothly before the saddle-node bifurcation at $Ra = 6157.97$. Prior to this, PO1 loses stability by undergoing a period-doubling bifurcation at $Ra=6154.7$ to PO2, which will be discussed in \S \ref{hetero_creation}. The saddle-node bifurcation can be seen in both the maximum and minimum dashed green curves of figure \ref{largedomain_BDall}(a) and leads to what we call the lower branch (because of its lower value of $\lvert\lvert \theta \lvert\lvert_2$). 

\par By using the multi-shooting method with two to five shots, we have been able to continue the lower PO1 branch down in Rayleigh number to $Ra=6152.2041$, where the period of PO1 is very long: $T=955.4$ time units. We will see below that PO1 disappears via a homoclinic bifurcation, at which its period is infinite. Figures \ref{global_bifurcation}(a)-(d) show snapshots of PO1 at $Ra = 6152.249$, on the lower branch. Among these snapshots, figures \ref{global_bifurcation}(a) and \ref{global_bifurcation}(b) capture the thinning and thickening of the rolls along the diagonal, with local waviness along the edge of the rolls. The waviness becomes weaker in figure \ref{global_bifurcation}(c) and finally in \ref{global_bifurcation}(d) the edges are smoother and the roll widths almost uniform. All of the states in the cycle have a definite diagonal orientation. This implies that there exists another version of PO1 with the opposite diagonal orientation. The times at which these snapshots are taken are marked by stars in figures \ref{global_bifurcation}(e-f). 

\par Figure \ref{global_bifurcation}(e) shows time series initialized with this unstable PO1 and also with FP4 (with a small perturbation), at $Ra=6152.249$. Both of these runs eventually converge to another state: the stable upper branch of PO1, whose period $T = 161$ is much shorter than the period $T=900$ of the lower branch PO1. For $t< 1000$, the red curve remains close to FP4 during a large portion of the period. Figure \ref{global_bifurcation}(d) corresponds to the fourth star of \ref{global_bifurcation}(e), indicating via this projection that \ref{global_bifurcation}(d) is long-lived and very close to FP4. Indeed we used figure \ref{global_bifurcation}(d) as the initial estimate for Newton's method to converge to FP4 at $Ra=6152.249$. However, figure \ref{global_bifurcation}(c), which only shows a transient at $Ra=6152.249$, resembles figure \ref{FP918}(e), which shows the converged FP4 at $Ra=6281$. We see from this that the diagonal orientation of FP4 becomes more prominent at higher Rayleigh numbers. Figure \ref{global_bifurcation}(f) shows a phase portrait visualization from the same simulation as \ref{global_bifurcation}(e), using the thermal energy input $I$ and viscous dissipation $D$. There, FP4 is shown as the hollow blue circle on the $D/I=1$ vertical line, showing that energy dissipation and input are equal. Near FP4, the dotted red curve looks continuous; this is due to the clustering of points near FP4. 

\begin{figure}
    \centering
    \includegraphics[width=\columnwidth]{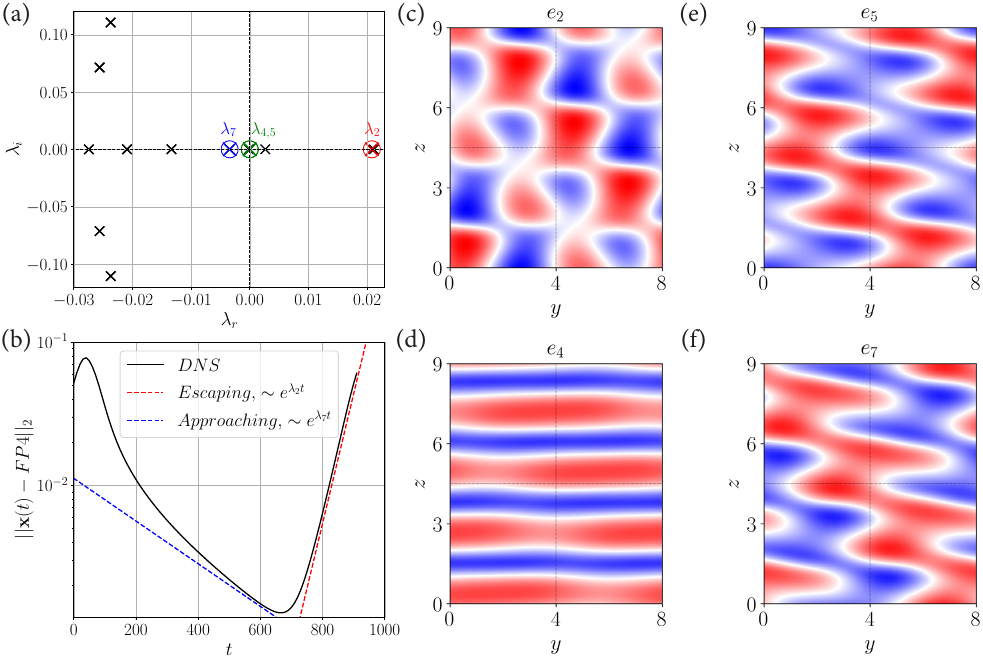}
    \captionsetup{font={footnotesize}}
    \captionsetup{width=13.5cm}
    \captionsetup{format=plain, justification=justified}       
    \caption{\label{approach} (a) Leading eigenvalues at $Ra=6152.249$ of FP4: $[\lambda_1, \lambda_2, \lambda_3, \lambda_4, \lambda_5, \lambda_6, \lambda_7] = [0.0212, 0.0208, 0.0026, 0, 0, -0.00017, -0.0034]$. Eigenvalues $\lambda_2$ (escaping, red), $\lambda_{4,5}$ (neutral, green) and $\lambda_7$ (approaching, blue) are marked in color. (b) $L_2$-distance between each instantaneous flow field of PO1 and FP4 at $Ra=6152.249$, close to $Ra_{\rm hom}=6151.97$. The evolution of PO1 (black curve) is exponential most of the time, with the escape from (red line) and approach to (blue line) FP4 governed by $\lambda_2$ and $\lambda_7$. (c-f) Four leading eigenmodes of FP4 at $Ra = 6152.249$, visualized via the temperature field on the $y$-$z$ plane at $x=0$: $e_2$, $e_{4}$, $e_{5}$ and $e_7$. The same color bar is used in all plots.}
\end{figure}

\subsubsection{Termination of PO1: homoclinic bifurcation}
\label{homo_termination}
\par The close approach to FP4 implies that PO1 is close to a homoclinic cycle. We have verified that this closest approach is always to the same version of FP4 and not to another symmetry-related version. Thus, PO1 approaches a homoclinic, and not a heteroclinic cycle. A homoclinic cycle approaches a fixed point along one of its stable directions and escapes from it along one of its unstable directions. For this reason, we compute the eigenvalues and eigenvectors of FP4. Figure \ref{approach}(a) shows the leading eigenvalues that we have computed at $Ra = 6152.249$, close to the global bifurcation point. The seven leading eigenvalues, all real, are $[\lambda_1, \lambda_2, \lambda_3, \lambda_4, \lambda_5, \lambda_6, \lambda_7] = [0.0212, 0.0208, 0.0026, 0, 0, -0.00017, -0.0034]$. We have set any eigenvalue whose absolute value is less than $10^{-7}$ to zero. Figure \ref{approach}(a) shows other eigenvalues with smaller real parts as well and some of the eigenvalues are too close together to be distinguished. Certain eigenvalues of special significance are highlighted by colored circles and their corresponding eigenvectors are shown in figures \ref{approach}(c-f). 

\par The eigenvectors can be interpreted by considering FP4 and PO1, as depicted in figures \ref{global_bifurcation}(a-d). There are two neutral directions, due to the continuous translation symmetry in the periodic directions. Eigenvalue $\lambda_4$ is zero and the corresponding eigenvector $e_4$, depicted in figure \ref{approach}(d), is the neutral mode associated with $z$-translation (i.e. the $z$ derivative) of the roll-like FP4, very close to what is depicted in figure \ref{global_bifurcation}(d). There must also be a marginal eigenvector corresponding to $y$-translation and indeed, $\lambda_5=0$ and we have verified numerically that $e_5$, depicted in figure \ref{approach}(e), is the $y$ derivative of FP4. This is not immediately obvious, but note that for $z$ constant, the $y$ derivative of FP4 oscillates in sign and its maxima and minima are located along a diagonal. The green circle in figure \ref{approach}(a) contains $\lambda_4$ and $\lambda_5$ (but also $\lambda_6$, whose decay rate is very small). 

\par The other two eigenvectors shown in figures \ref{approach}(c) and (f) are responsible for the approach to and escape from FP4. We have determined which eigenvalues are associated with approach and escape by comparing them with the observed approach and escape rates, and also by subtracting FP4 from the instantaneous flow fields and comparing the result to the eigenvectors. For the escaping dynamics of PO1 from FP4, the quantity ($||\boldsymbol{x}(t) - \text{FP4}||_2$) increases exponentially at rate $\lambda_2=0.0208$. The corresponding eigenvector $e_2$ is shown in figure \ref{approach}(c) and can be viewed as corresponding to widening and narrowing of the rolls. The approaching dynamics is characterized by $\lambda_7=-0.0034$. The corresponding eigenvector $e_7$, shown in figure \ref{approach}(f), can be viewed as corresponding to translation in $y$. The portion of PO1 escaping FP4 along $e_2$ can be seen as the red line in figure \ref{approach}(b); the escaping portion is fit to the red line. While the rate of escape matches $\lambda_2$ closely, the approach rate only fits $\lambda_7$ over a short range of time. In figure \ref{approach}(a), the red circle contains $\lambda_2$ (but also $\lambda_1$, which is very close to $\lambda_2$), while the blue circle encloses $\lambda_7$. 

\par The stable eigendirection $e_7$ along which PO1 approaches FP4 is not the leading stable (least negative) one, as would be usual for a homoclinic orbit. This is because PO1 exists in the invariant symmetry-restricted subspace $\braket{\tau(L_y/4,-L_z/4)}$, to which $e_7$ also belongs. In contrast, $e_6$ (not shown) has the opposite symmetry $\braket{\tau(L_y/4,L_z/4)}$. Note also that near-homoclinic orbits for which the rate of escape exceeds the rate of approach (i.e. here $|\lambda_2| >|\lambda_7|$) are unstable, as is already seen in the time series in figure \ref{global_bifurcation}(e). However, because our periodic orbits are computed using Newton's method and not time integration, we can calculate this periodic orbit despite its instability.

\par In addition, a homoclinic orbit bifurcating from a hyperbolic fixed point (which is the case for FP4) is structurally unstable, i.e.\ it exists for a single parameter value; see \citet[Lemma 6.1]{Kuznetsov2004} for a proof. Strictly speaking, FP4 is a relative hyperbolic fixed point, since it has zero eigenvalues along the directions of its continuous translation symmetries in $y$ and $z$, but the result applies to the evolution normal to these directions, i.e. with $y$ and $z$ phases fixed \citep{krupa1995asymptotic}. Thus, the homoclinic cycle on which PO1 terminates is neither stable nor robust.

\par The closeness of some of the eigenvalues in figure \ref{approach}(a) can be explained by the fact that the $y$ dependence of FP4 is extremely weak. If FP4 were entirely $y$-independent, like FP1, then eigenvectors would come in pairs, corresponding to a trigonometric dependence (analogous to $\rm sine$- and $\rm cosine$-eigenmodes) in $y$ with different phases, or to the choice of diagonal direction. Since the dependence in $y$ is weak, this is still approximately true in many cases. Eigenvalue $\lambda_1=0.0212$ is very close to $\lambda_2=0.0208$ and indeed eigenvector $e_1$ (not shown) resembles a $y$-shifted version of $e_2$. The near-neutral eigenvalues $\lambda_3=0.0026$ and $\lambda_6=-0.00017$ correspond to eigenvectors $e_3$ and $e_6$ (not shown), which resemble $e_5$ and $e_7$ but oriented in the opposite diagonal direction or, equivalently, reflected. 

\begin{figure}
    \centering
    \includegraphics[width=\columnwidth]{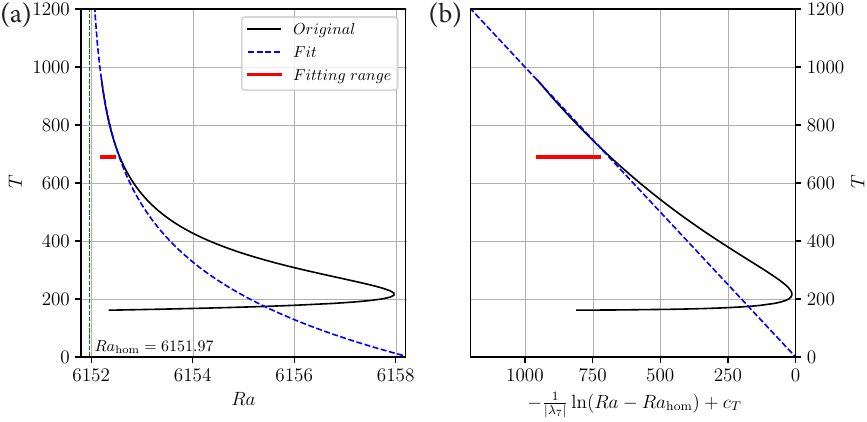}
    \captionsetup{font={footnotesize}}
    \captionsetup{width=13.5cm}
    \captionsetup{format=plain, justification=justified} 
    \caption{\label{Period_div_PO1} Growth of the period of PO1 close to the global bifurcation point. PO1 undergoes a saddle-node bifurcation at $Ra = 6157.97$ where the lower branch appears. (a) shows the periods computed by numerical continuation and its logarithmic fit (see text). (b) uses a logarithmic scale for $Ra-Ra_{\rm hom}$, on which the period depends linearly. The red horizontal bar in (a) and (b) indicates the Rayleigh number range $6152.2<Ra<6152.45$ used for curve fitting.}
\end{figure}

\par As $Ra$ approaches $Ra_{\rm hom}$, PO1 approaches FP4 and the time spent near FP4 increases, until PO1 touches FP4 and acquires an infinite period in a homoclinic bifurcation. The period of PO1 is dominated by the time of approach to FP4, as shown in figure \ref{approach}(b). This time can be estimated by the formula 
\begin{equation}
    T \approx -\frac{1}{|\lambda_-|}\ln|Ra-Ra_{\rm hom}| + c_T,
\end{equation}
where $\lambda_-=\lambda_7=-0.0034$ is the rate of exponential approach to FP4, $Ra_{\rm hom} = 6151.97$ and $c_T=533$ is a fitting constant. This asymptotic scaling law for $T$ as a function of $Ra$ was first derived in \citet{Gaspard1990} and later used by various researchers including \citet{Meca2004, Reetz2020b, Liu_Sharma_Julien_Knobloch_2024}. As shown in figure \ref{Period_div_PO1}, we have fit the numerically computed periods of the states on the lower branch to this formula. Note that only the Rayleigh number range $6152.2<Ra<6152.45$ very close to $Ra_{\rm hom}$ has been used for fitting and that we have extended the backward continuation of PO1 to the lowest Rayleigh number possible ($Ra=6152.2041$) within our numerical precision and ability. 

\par \citet{Gao2018} observed a periodic orbit produced by a Hopf bifurcation from a steady state; these are the solutions that we have called PO1 and FP3. Our bifurcation analysis agrees with their results. Extending their work, we have found that PO1 undergoes a saddle-node bifurcation and then terminates in a homoclinic bifurcation by meeting a new unstable fixed point, FP4.

\subsection{Second periodic orbit (PO2)}
\subsubsection{Creation of PO2: period-doubling bifurcation and symmetry}
\label{hetero_creation}

\begin{figure}
    \centering
    \includegraphics[width=\columnwidth]{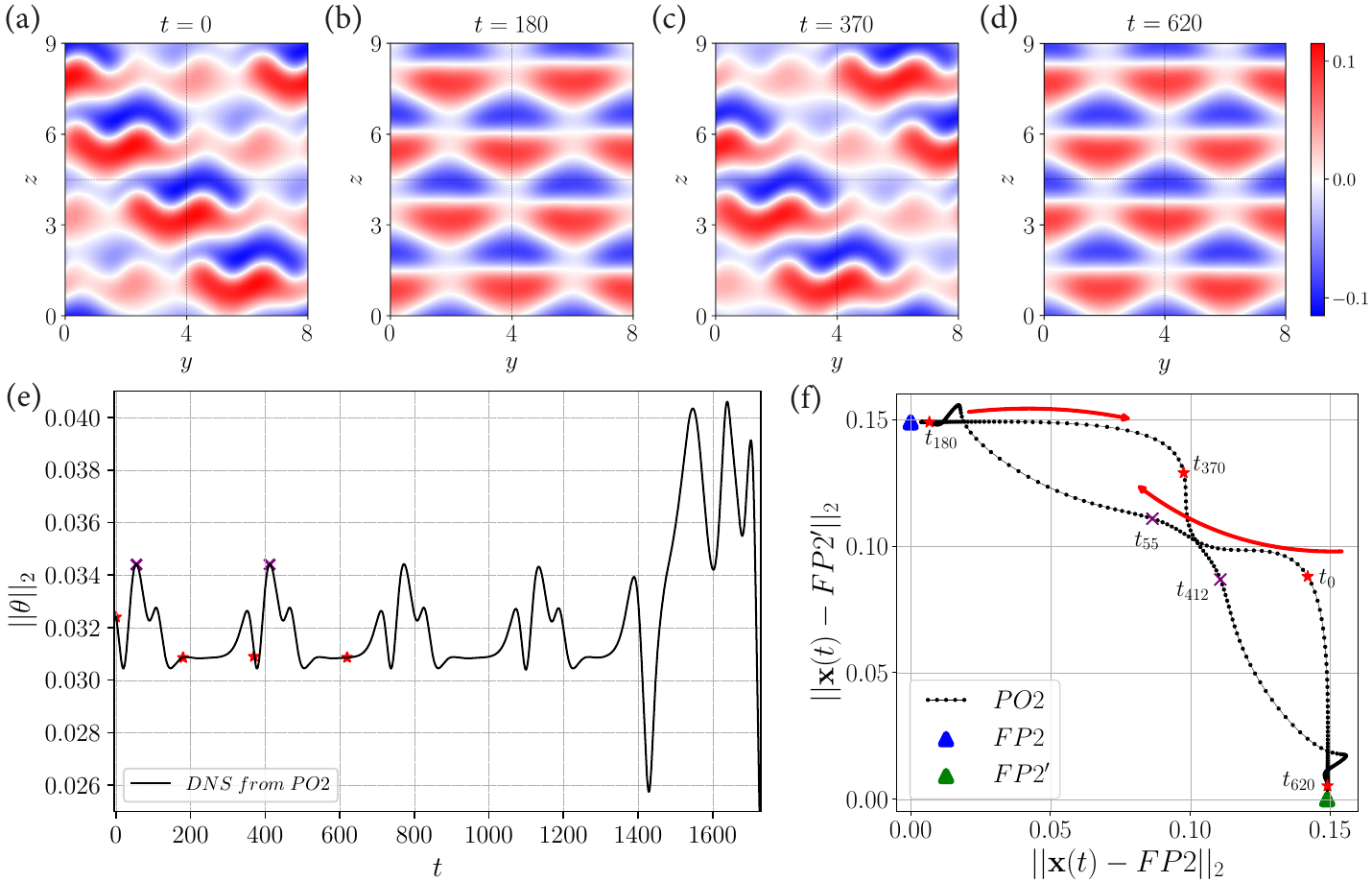}
    \captionsetup{font={footnotesize}}
    \captionsetup{width=13.5cm}
    \captionsetup{format=plain, justification=justified}
    \caption{\label{PO2_hetero} (a-d) Snapshots of the dynamics of PO2 (visualized via the temperature field on the $y$-$z$ plane at $x=0$) at $Ra=6277.88$ near $Ra_{\rm het} = 6277.95$. Snapshots (b) and (d) show states which are close to two symmetry-related versions of FP2 (figure \ref{FP918}(c)). (e) Time series from DNS at $Ra=6277.88$, initialized by the unstable PO2 shown in (a). The dynamics after $t\approx1250$ becomes irregular and eventually terminates in chaos. (f) Phase space projection close to the global bifurcation point: shown are the PO2 at $Ra=6277.88$ and two symmetry-related FP2 states involved in the heteroclinic cycle. In (e) and (f), the four red stars indicate the moments where the snapshots (a)-(d) are taken and the two purple crosses mark the instants $t_{55}$ and $t_{412}$. In (f), the red arrows show the direction of the trajectory.}
\end{figure}

\par PO2 bifurcates from PO1 in a period-doubling bifurcation at $Ra=6154.7$. At this value of Rayleigh number, its period ($T=341$) is exactly twice that of PO1 ($T=170.5$), as shown in figure \ref{largedomain_BDall}(b). We have confirmed the threshold in two additional ways: at $Ra=6154.7$, the maxima and minima of $\lvert\lvert \theta \lvert\lvert_2$ of PO2 in the time series are extremely close in amplitude and frequency to those of PO1, and the power spectrum contains a very small component of the new frequency of PO2. 

\par PO2 inherits all of the spatial symmetries of PO1: $\braket{\pi_y\pi_{xz}, \tau(L_y/4,-L_z/4)} \simeq D_4$ and has in addition the spatio-temporal symmetry:
\begin{equation}
\label{RPO}
(u,v,w,\theta)(x,y,z,t+T/2) = (u,v,w,\theta)(x,y-L_y/4,z,t).
\end{equation}
\cite{Gao2018} presented visualisations of PO2 in their figure 20 at $Ra=6250$ and noted that it satisfied \eqref{RPO}. Figures \ref{PO2_hetero}(a-d) show four snapshots of the temperature field of PO2 in which the spatio-temporal symmetry \eqref{RPO} of PO2 can clearly be seen.  Figures \ref{PO2_hetero}(b) and (d) are very similar to each other and to two symmetry-related versions of the diamond-roll state, which we denote by FP2 and FP2$^\prime \equiv \tau(L_y/4=2,0)$FP2. Between these instants, figures \ref{PO2_hetero}(a) and (c) show a wavy modulation of convection rolls along one of the diagonals. Since FP2 is $y$-reflection symmetric, there necessarily exists another version of PO2 in which the modulation occurs along the other diagonal. 

\subsubsection{Termination of PO2: heteroclinic bifurcation and eigendirections}
\label{termination_PO2}
\begin{figure}
    \centering
    \includegraphics[width=\columnwidth]{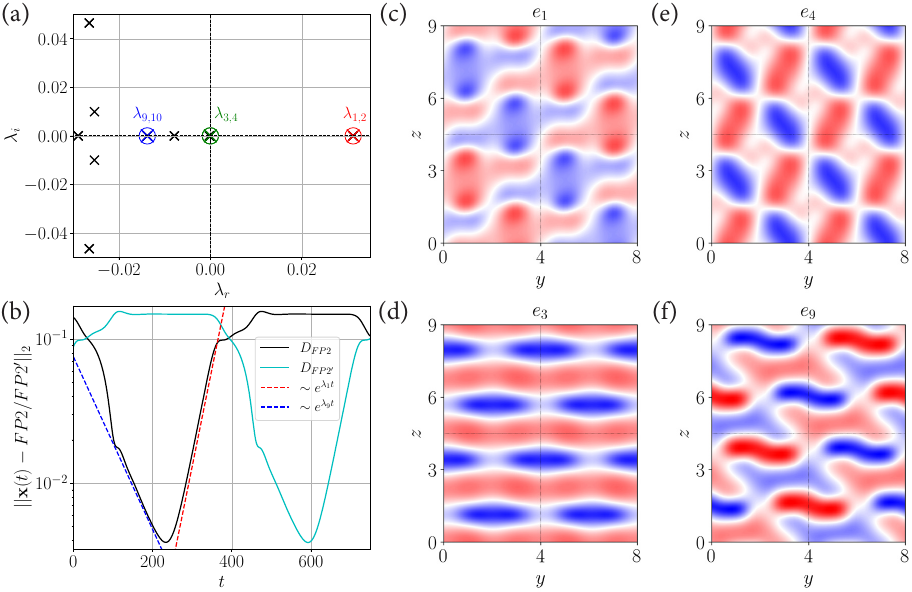}
    \captionsetup{font={footnotesize}}
    \captionsetup{width=13.5cm}
    \captionsetup{format=plain, justification=justified} 
    \caption{\label{approach_hetero} (a) Leading eigenvalues of FP2 at $Ra=6277.88$. The ten leading eigenvalues are real and double: $[\lambda_{1,2}, \lambda_{3,4}, \lambda_{5,6}, \lambda_{7,8}, \lambda_{9,10}]=[0.031, 0, -0.00019, -0.00788, -0.0138]$. Eigenvalue $\lambda_{1,2}$ (escaping, red), $\lambda_{3,4}$ (neutral, green) and $\lambda_{9,10}$ (approaching, blue) are marked in color. (b) $L_2$-distance between each instantaneous flow field of PO2 and FP2 (and FP2$^\prime$) at $Ra=6277.88$, close to $Ra_{\rm het}$. The dynamics of PO2 is exponential for most of the cycle (black and cyan curves). The approaching (blue line) and escaping (red line) dynamics of PO2 with respect to FP2 are shown and are governed by two eigenvalues of FP2. (c-f) Four leading eigenmodes of FP2 at $Ra=6277.88$, visualized via the temperature field on the $y$-$z$ plane at $x=0$: $e_1$, $e_3$, $e_4$ and $e_9$. The same color bar is used in all plots.}
\end{figure}

\par Figure \ref{largedomain_BDall}(b) shows that although the period of PO2 decreases significantly with increasing $Ra$ until $Ra=6204.8$, it increases beyond that, eventually diverging. This implies that PO2 undergoes a global bifurcation. We have been able to continue PO2 until $(Ra,T)=(6277.88, 710.8)$ and we estimate the critical Rayleigh number for the global bifurcation to be approximately $Ra_{\rm het} = 6277.95$. 

\par The snapshots of figures \ref{PO2_hetero}(a-d) are taken at $Ra=6277.88$, very close to $Ra_{\rm het} = 6277.95$. PO2's alternating visits to FP2 and FP2$^\prime$ indicate that PO2 ends at a heteroclinic cycle between these two fixed points. In figure \ref{PO2_hetero}(e), we show a time series of PO2 at $Ra = 6277.88$, indicating the instants at which snapshots in figures \ref{PO2_hetero}(a-d) are taken. It is clear that figures \ref{PO2_hetero}(b) and (d) belong to fairly long-lived plateaux. The global measurement $||\theta||_2$ does not distinguish between FP2 and FP2$^\prime$, so we have plotted a phase portrait in figure \ref{PO2_hetero}(f), which represents each instantaneous flow field by its distance from each version of FP2. The phase portrait shows the clustering of points near FP2 and FP2$^\prime$, confirming that PO2 is close to a heteroclinic cycle connecting these states. 

\par The phase portrait in figure \ref{PO2_hetero}(f) also shows clustering of points around $(0.1, 0.1)$, corresponding to instants $t_{38}$ and $t_{394}$. This clustering suggests that the limit cycle might be approaching other fixed points. However, the time series in figure \ref{PO2_hetero}(e) does not show any other plateaux close to $t_{38}$ and $t_{394}$, and Newton's method did not converge to any new equilibria around the states from $t_{28}$ to $t_{55}$ and from $t_{384}$ to $t_{412}$. This remains true up to the highest Rayleigh number (or equivalently, the longest period) of PO2 reached by our numerical continuation. We conclude that this heteroclinic cycle contains no other fixed points.
 
\par The dynamics along which PO2 approaches and escapes from FP2 can be described by eigenvalues and eigenvectors of FP2. Figure \ref{approach_hetero}(a) shows the leading eigenvalues of FP2 at $Ra = 6277.88$, computed by Arnoldi iteration, and which are $[\lambda_{1,2}, \lambda_{3,4}, \lambda_{5,6}, \lambda_{7,8}, \lambda_{9,10}] = [0.031, 0, -0.00019, -0.00788, -0.0138]$. We previously saw that for FP4, the eigenvalues are approximately double (see figure \ref{approach}(a)) due to the approximate symmetries of FP4. Here, FP2 has exact reflection symmetries leading to eigenvalues which are exactly double. 

\par The two neutral eigenmodes due to the continuous translation symmetries are $e_3$, corresponding to the $z$ derivative of FP2 and shown in figure \ref{approach_hetero}(d) and $e_4$, corresponding to its $y$ derivative and shown in figure \ref{approach_hetero}(e). The green circle in figure \ref{approach_hetero}(a) contains $\lambda_{3,4}$, but also $\lambda_{5,6}$. Figure \ref{approach_hetero}(c) shows the escaping eigenmode $e_1$, which is responsible for choosing the diagonal orientation of PO2. Looking at figures \ref{PO2_hetero}(a-d), this is not obvious, but we have verified that subtracting FP2 from instantaneous temperature fields in the escaping phase of PO2 yields a field resembling $e_1$. Moreover, eigenvalues $\lambda_{1,2}=0.031$ capture well the escape rate from FP2, as shown in figure \ref{approach_hetero}(b). Eigenmode $e_2$, with the same eigenvalue, is related to $e_1$ by reflection symmetry, as shown in figures \ref{D4_FP2}(a)-(b) for $Ra=6056$. The direction in which PO2 approaches FP2 is $e_9$, again confirmed by subtracting FP2 from the appropriate flow field in PO2, and $\lambda_{9,10}$ closely approximate the decay rate to FP2 shown in figure \ref{approach_hetero}(b). The direction in which PO2 approaches the equilibrium is again not its leading stable eigendirection, and for the same reason as for PO1: PO2 remains within the symmetry group $\braket{\pi_y\pi_{xz}, \tau(L_y/4,-L_z/4)}$, to which $e_9$ belongs, but not eigenmodes $e_{5,6}$ and $e_{7,8}$ (not shown). Since $|\lambda_{1,2}|>|\lambda_{9,10}|$, the heteroclinic cycle is unstable, which is confirmed by the chaotic behavior in the time series in figure \ref{PO2_hetero}(e) after $t\approx1250$. 

\par For FP2, since the eigenvalues are double, the eigenspace corresponding to each is two-dimensional; the eigenvectors that play the roles mentioned above -- $y$-translation, $z$-translation, escape, and approach -- must be selected as linear combinations of the two arbitrary eigenvectors returned by the Arnoldi method. By differentiating FP2 in $y$ and $z$ and by subtracting FP2 from the instantaneous flow fields during approaching or escaping phases, we have been able to choose the appropriate eigenvector in each case.

\subsubsection{Robust heteroclinic cycle and 1:2 resonance}
\label{robust_hetero_cycle}
\par
\begin{figure}
    \centering
    \includegraphics[width=\columnwidth]{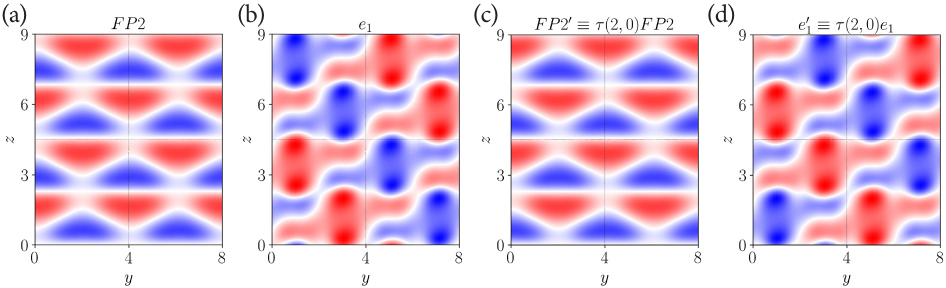}
    \captionsetup{font={footnotesize}}
    \captionsetup{width=13.5cm}
    \captionsetup{format=plain, justification=justified} 
    \caption{\label{eigen_heteroFP2} (a) and (c) FP2 and FP2$^\prime\equiv\tau(2,0)$FP2 at $Ra=6277.8$. (b) and (d) Unstable eigenmodes: $e_1$ of FP2 and $e_1^\prime\equiv\tau(2,0)e_1$ of FP2$^\prime$. The wavenumbers of the equilibria and unstable eigenmodes in the $y$ direction suggest a 1:2 mode interaction. All snapshots are visualized via the temperature field on the $y$-$z$ plane at $x=0$.}
\end{figure}

 We now wish to show that the heteroclinic cycle that PO2 approaches is robust (also called structurally stable) i.e.\ that it exists over a parameter range rather than only at a single point. We have confirmed by numerical experiments that varying slightly the Rayleigh number does not affect the two transitions, also called half-cycles, FP2$\rightarrow$FP2$^\prime$ and FP2$^\prime$$\rightarrow$FP2. More rigorously, we list here the three conditions \citep{Krupa1997} that are required for a heteroclinic cycle between two fixed points, here FP2 and FP2$^\prime$, to be robust:
\begin{enumerate}
    \item There exist two invariant subspaces $S$ and $S^\prime$ such that FP2 is a saddle (sink) and FP2$^\prime$ is a sink (saddle) for the flow restricted to subspace $S$ ($S^\prime$). \label{itemi}
    \item There exist saddle-sink connections FP2$\rightarrow$FP2$^\prime$ in $S$ and FP2$^\prime$$\rightarrow$FP2 in $S^\prime$.\label{itemii}
    \item There exists a symmetry relation between the two fixed points. \label{itemiii}
\end{enumerate}
Item \eqref{itemiii} is satisfied by definition: we have set FP2$^\prime \equiv \tau(L_y/4,0)$FP2. For items \eqref{itemi} and \eqref{itemii}, we define $S$ and $S^\prime$ to be the fixed-point subspaces of two conjugate subgroups:
\begin{equation}
\begin{array}{ll}
 S &\equiv {\rm Fix}|_{\braket{\pi_y\pi_{xz}\tau(L_y/2,0), \tau(L_y/4,-L_z/4)}},\\
 S^\prime &\equiv {\rm Fix}|_{\braket{\pi_y\pi_{xz}, \tau(L_y/4,-L_z/4)}}.
\end{array}
\end{equation}
We note that $e_1$, depicted in figure \ref{eigen_heteroFP2}(b), is an unstable eigenvector of both FP2 and FP2$^\prime$ and belongs to subspace $S$ but not to $S^\prime$. We define $e_1^\prime \equiv \tau(L_y/4,0)e_1$, shown in figure \ref{eigen_heteroFP2}(d), which is also an unstable eigenvector of FP2 and FP2$^\prime$, and which belongs to subspace $S^\prime$ but not to $S$. The $L_2$-inner product of these two eigenmodes $\langle e_1, e_1^\prime \rangle$ is zero, and so they are orthogonal.

\par We have carried out simulations within subspaces $S$ and $S^\prime$ by numerically imposing the corresponding symmetries. When we restrict the simulation to $S$, the unstable eigenmode $e_1^\prime$ is disallowed and escape from FP2 (a saddle in subspace $S$) must take place along $e_1$. This trajectory lands on FP2$^\prime$, which is linearly stable (a sink in subspace $S$). Changing the imposed subspace from $S$ to $S^\prime$, eigenvector $e_1$ is disallowed, and escape from FP2$^\prime$ (a saddle in subspace $S^{\prime}$) occurs along $e_1^\prime$. This trajectory lands on the stable equilibrium FP2, which is a sink in subspace $S^{\prime}$. Similar arguments apply to the approaches to FP2$^\prime$ and FP2 via eigenvectors $e_9^\prime$ and $e_9$, shown in figure \ref{approach_hetero}(f), respectively. Thus, we have demonstrated items \eqref{itemi} and \eqref{itemii}, proving that the heteroclinic cycle is robust. These three conditions are also discussed in \citet{Reetz2020a}, together with an example of a robust heteroclinic cycle between two symmetrically-related oblique-wavy-roll equilibrium states found in inclined layer convection system.

\par In addition to demonstrating that the heteroclinic cycle that emerges from PO2 and FP2 is robust, we will discuss its origin. We first address why FP2 has unstable and stable eigenvectors of the form $e_1$ and $e_9$. We recall that FP1 is homogeneous in $y$, FP2 has a $y$ periodicity of $L_y/2=4$, and FP3, FP4, and FP5 have $y$-periodicity $L_y=8$. When FP2 is created, it inherits the eigenvectors of FP1, including those which lead from FP1 to FP4 and FP5 (with $y$-periodicity $L_y=8$). The existence of such eigenvectors for FP2 is confirmed by the bifurcations from it to FP3 and FP5, which also have $y$-periodicity $L_y=8$; see, for example figure \ref{D4_FP2}. Because these $L_y$-periodic eigenvectors are all associated with nearby bifurcations, their corresponding eigenvalues are necessarily among the leading ones of FP2 in this range of $Ra$. This is the scenario of 1:2 resonance, the normal form of which was derived by \cite{Armbruster1988}:
\begin{equation}
\begin{array}{ll}
\dot{z_1} &= \overline{z_1} z_2 + z_1 \left(\mu_1 + e_{11} |z_1|^2 + e_{12} |z_2|^2 \right), \\
\dot{z_2} &= \pm z_1^2 + z_2 \left(\mu_2 + e_{21} |z_1|^2 + e_{22} |z_2|^2 \right), 
\end{array}
\label{eq:sysz}
\end{equation}
where $z_1$ and $z_2$ are complex amplitudes of modes with wavenumbers 1 and 2, $\mu_1, \mu_2$ are control parameters and $e_{11}, e_{12}, e_{21}, e_{22}$ are nonlinear coefficients. These authors have demonstrated that \eqref{eq:sysz} has a solution which is a heteroclinic orbit over a finite range of parameter values. Heteroclinic orbits of this type have been observed in full fluid dynamical configurations by, e.g.\ \cite{Mercader2002, Nore2003mode, Reetz2020a}.

\par We now recall from \S \ref{two_simu_bif_part2} that, in addition to the diagonally-oriented eigenmode $e_1$ and its translation- and reflection-related versions, FP2 also has eigenmodes of type $e_2\equiv (e_1+\pi_y e_1)/\sqrt{2}$, shown in figure \ref{D4_FP2}(c,d,e,f), which have a reflection symmetry in $y$ and which we have called rectangular. The diagonal eigenvector $e_1$ is responsible for the bifurcation to FP3, while the rectangular eigenvector $e_2$ is responsible for the  bifurcation to FP5. Perturbing FP2 along $e_2$ can lead to a rectangular periodic orbit that retains $y$-reflection symmetry, which is currently under investigation.

\subsubsection{Stability of PO2}
\label{sec:stabilityPO2}
\begin{figure}
    \centering
    \includegraphics[width=\columnwidth]{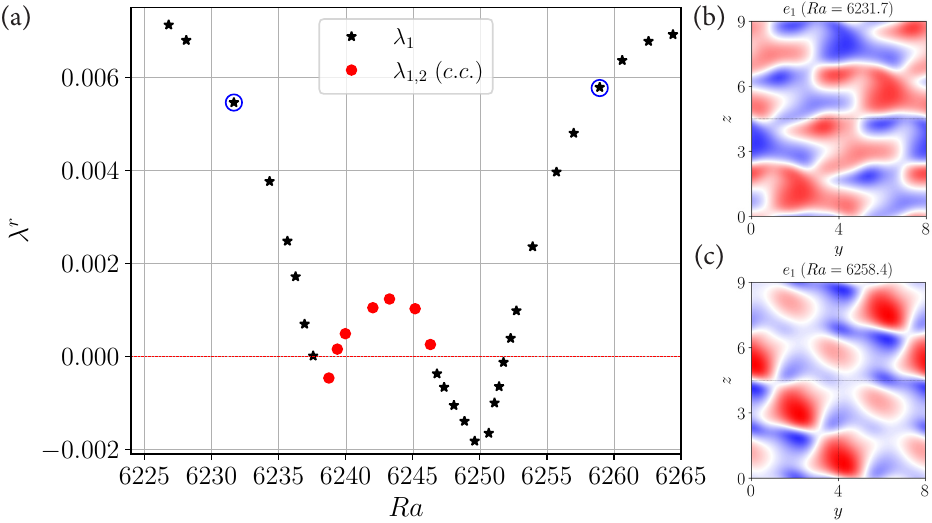}
    \captionsetup{font={footnotesize}}
    \captionsetup{width=13.5cm}
    \captionsetup{format=plain, justification=justified}
    \caption{\label{evo_eigen_PO2} (a) The real part of the leading Floquet exponents of PO2 as a function of Rayleigh number. From low to high Rayleigh number, the leading Floquet exponent decreases monotonically within $6226<Ra<6238.75$. At $Ra=6238.75$, one sees the formation of a complex conjugate pair which has a positive real part for $6239<Ra<6246.32$. For $6246.5<Ra<6252$, PO2 is stable, with stability lost for $Ra>6252$. The apparent non-smoothness of the curve at $(Ra,\lambda_1)\approx (6238.7,-0.00047)$ and $(6250,-0.002)$ is due to the crossover of competing leading Floquet exponents. The two blue circles indicate where (b) and (c) are taken. (b-c) Two leading unstable Floquet eigenmodes for $6226<Ra<6237.6$ (b) and for $Ra>6252$ (c), visualized via the temperature field on the $y$-$z$ plane at $x=0$. The same color bar is used in (b) and (c).}
\end{figure}

\par PO2 is stable over a short interval: from its onset at $Ra=6154.7$ until $Ra=6173.8$, where it becomes unstable via a pitchfork bifurcation giving rise to another periodic orbit PO3, to be discussed next in \S \ref{PO3}. Just before the global bifurcation at $Ra=6277.95$, PO2 undergoes two saddle-node bifurcations at $Ra=6276$ and then at $Ra=6273.6$; these bifurcations do not restabilize PO2. However, \cite{Gao2018} observed PO2 at $Ra=6250$ via DNS, implying that PO2 should be stable at that Rayleigh number. In order to understand this, we computed the leading Floquet exponent of PO2 over a range of $Ra$ surrounding 6250. 

\par The intriguing evolution of the stability of PO2 is presented in figure \ref{evo_eigen_PO2}. The leading Floquet exponent $\lambda_1$ is real from $Ra=6173.8$ to $Ra=6237.6$: it increases monotonically from $Ra=6173.8$ to $Ra=6225$ (not shown), and then decreases monotonically to zero over the interval $6226<Ra<6237.6$. The leading real exponent is then superseded by a complex conjugate pair $\lambda_{1,2}$ whose real part, initially negative, becomes positive over the interval $6239<Ra<6246.32$. The leading exponent $\lambda_1$ is then real and negative, so that there is a small interval $6246.5<Ra<6252$ over which PO2 is stable. It is within this very short interval that PO2 was observed by \citet{Gao2018}. In a further effort to understand the stabilization and subsequent destabilization of PO2 in this region, we computed the Floquet eigenmode to the left (figure \ref{evo_eigen_PO2}(b)) and right (figure \ref{evo_eigen_PO2}(c)) of the stable region, but we were unsuccessful in gleaning any physical insight from these. (There necessarily exist new branches bifurcating at the values at which $\lambda_1$ or the real part of $\lambda_{1,2}$ cross zero, but finding and following these new branches are beyond the scope of the current work.) 

\subsection{Third periodic orbit (PO3): pitchfork bifurcation}
\label{PO3}
\begin{figure}
    \centering
    \includegraphics[width=\columnwidth]{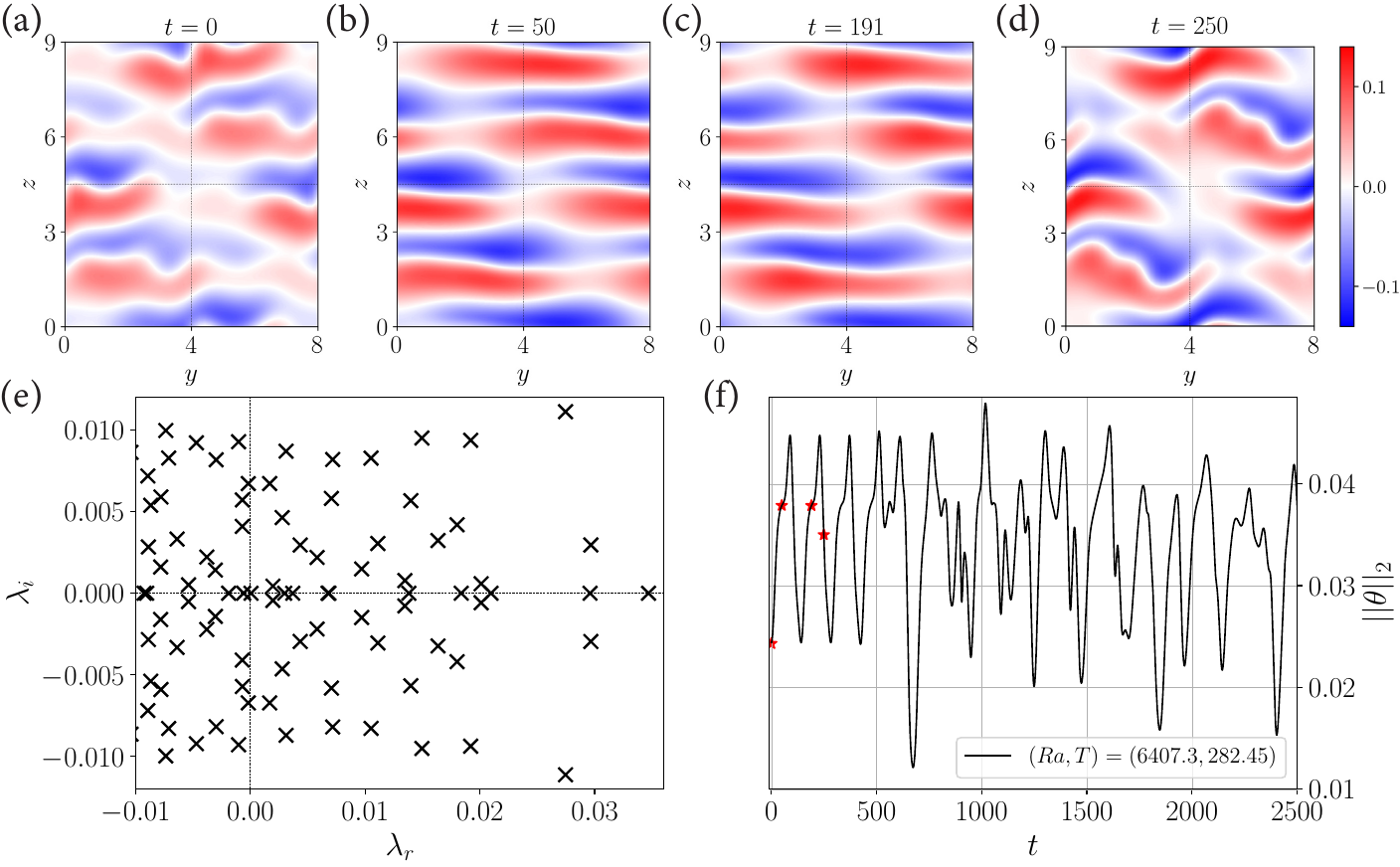}
    \captionsetup{font={footnotesize}}
    \captionsetup{width=13.5cm}
    \captionsetup{format=plain, justification=justified}
    \caption{\label{PO3-DNS} (a-d) Snapshots of the dynamics of PO3 (visualized via the temperature field on the $y$-$z$ plane at $x=0$) at $Ra=6407.3$ showing turbulent and disordered switching rolls. (e) Floquet exponent spectrum of PO3 at $Ra=6407.3$ showing its 51 unstable Floquet exponents. (f) Time series from DNS at $Ra =6407.3$, initialized by the converged unstable PO3. The temporal transition from a periodic to chaotic signal occurs at $t\approx400$. The red stars indicate the moments at which the snapshots (a)-(d) are taken.} 
\end{figure}

\par As mentioned in \S \ref{sec:stabilityPO2}, PO2 loses stability at $Ra=6173.8$ via a supercritical pitchfork bifurcation which creates PO3. The visual features of PO3 resemble those of PO2 near onset, but become much less regular at higher Rayleigh numbers, for instance at $Ra=6407.3$, depicted in figures \ref{PO3-DNS}(a) and (d). PO3 has spatial symmetries $\braket{\tau(L_y/2,-L_z/2)} \simeq Z_2$, and the spatio-temporal symmetry \eqref{RPO} inherited from PO2. This spatio-temporal symmetry can be seen by comparing figures \ref{PO3-DNS}(b) and (c), for instance; the direction of drift for PO3 is from left to right. PO3 loses stability at $Ra = 6183$. The bifurcating Floquet exponent is real, suggesting a pitchfork bifurcation leading to the creation of a pair of symmetry-related periodic orbits. However, we did not find any stable periodic orbit via DNS in the vicinity of $Ra=6183$, implying that such a bifurcation would be subcritical. Because PO3 is only stable for $6173.8 \leq Ra \leq 6183$, it is not surprising that it was not observed by \citet{Gao2018}. 

\par We continued PO3 until $Ra = 6407.3$, considerably into the chaotic regime ($Ra>6300$) mentioned by \citet{Gao2018}. (The range $6340<Ra<6407.3$ is not included in figure \ref{largedomain_BDall}.) Parametric continuation of PO3 for $Ra>6350$ was computationally challenging, probably due to the fact that the orbit is very unstable in this Rayleigh number range; see discussion of the numerical convergence of the iterative Newton algorithm in \citet{Sanchez2004} and \citet{Reetz2020b}. The spectrum of PO3 at $Ra=6407.3$ has more than $50$ unstable eigendirections with a wide range of frequencies, as illustrated in figure \ref{PO3-DNS}(e). Moreover, integrating the converged PO3 forward in time at $Ra = 6407.3$, the transition from a periodic to chaotic state is triggered after fewer than two periods of the orbit; see figure \ref{PO3-DNS}(f). Consequently, we stopped the forward Rayleigh number continuation at $Ra = 6407.3$ and do not discuss how PO3 terminates.

\subsection{Fourth periodic orbit (PO4): Hopf bifurcations}
\label{PO4_Hopf}
\begin{figure}
    \centering
    \includegraphics[width=\columnwidth]{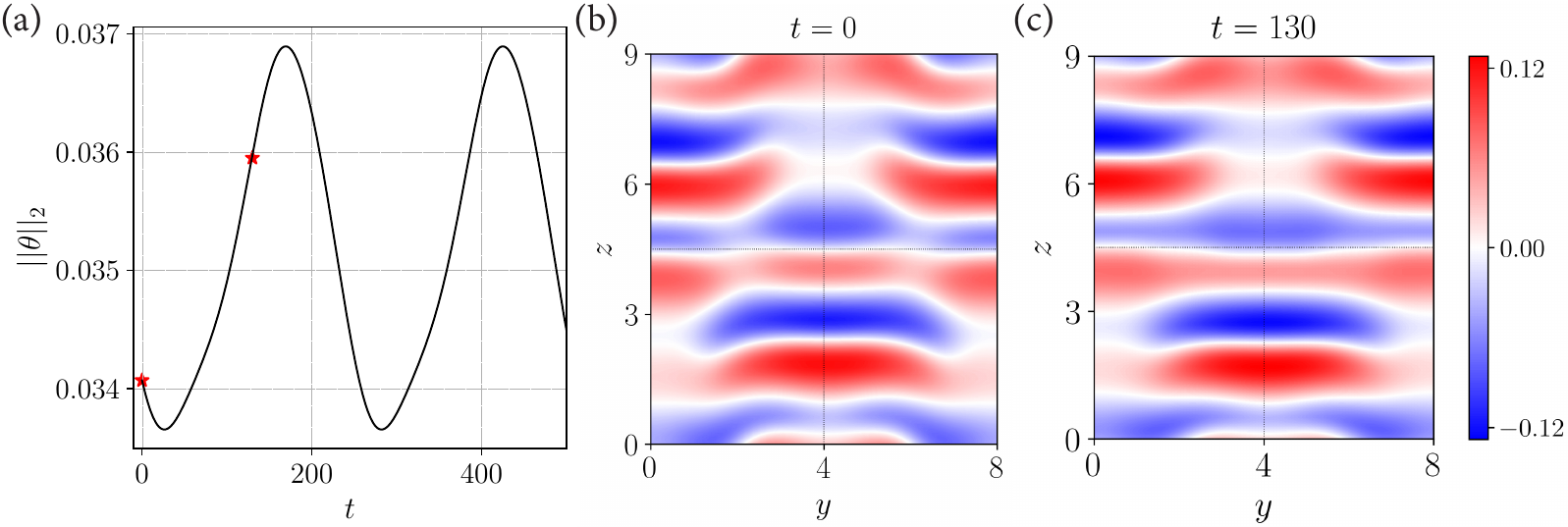}
    \captionsetup{font={footnotesize}}
    \captionsetup{width=13.5cm}
    \captionsetup{format=plain, justification=justified}
    \caption{\label{PO4-DNS} (a) Time series from DNS of PO4 at $Ra =6300$ ($T=255.7$). The red stars indicate the moments at which the snapshots (b) and (c) are taken. (b-c) Visualisations of PO4 at $Ra=6300$, via the temperature field on the $y$-$z$ plane at $x=0$.} 
\end{figure}

\par A new periodic orbit PO4 begins and ends on the lower branch of FP6 via two Hopf bifurcations at $Ra=6257.6$ and $Ra=6328.8$, respectively. As might be expected and as shown in figure \ref{PO4-DNS}, PO4 is an oscillating version of FP6. Since PO4 preserves the two reflection symmetries of FP6 $\braket{\pi_y,\pi_{xz}\tau(L_y/2,0)}$, PO4 has no additional spatio-temporal symmetries. The Hopf bifurcation terminating PO4 occurs very slightly before the saddle-node bifurcation that terminates FP6 at $Ra=6329$. At $Ra=6278$, PO4 is destabilized by the occurrence of a secondary Hopf bifurcation. Thus, PO4 is stable for $6257.6<Ra<6278$, as shown in figure \ref{largedomain_BDall}(a) and in the schematic figure \ref{Schematic-BD}. Its period increases smoothly and monotonically throughout its range of existence, shown in figure \ref{largedomain_BDall}(b).

\par Based on the bifurcation diagram in figure \ref{largedomain_BDall}(a), the family of branches FP5, FP6, and PO4, are unusual in leaving no trace of their existence beyond the disappearance of FP6 at $Ra=6329$. Two FP5 branches join and terminate at $Ra=6317.5$; two FP6 branches, themselves created from FP5, annihilate at $Ra=6329$; PO4, created from FP6, disappears at $Ra=6328.8$. When we add to this the disappearances of periodic orbits (PO1 and PO2) via global bifurcations, we see that of the six fixed points and four periodic orbits that arise from the bifurcation of 2D rolls at $Ra=5707$, only four fixed points and one periodic orbit survive past $Ra=6329$. \cite{Clever1995} comment about simplification in another phenomenon (drifting waves) that they observed in vertical convection: ``Of course, this is not a physically realistic scenario since there are other bifurcation points on the branch of the steady solutions ... But the return from a complex structure to a more simple one with increasing control parameter is a possibility that cannot be excluded a priori.''

\section{Discussion, conclusions and outlook}
\begin{figure}
    \centering
    \includegraphics[width=\columnwidth]{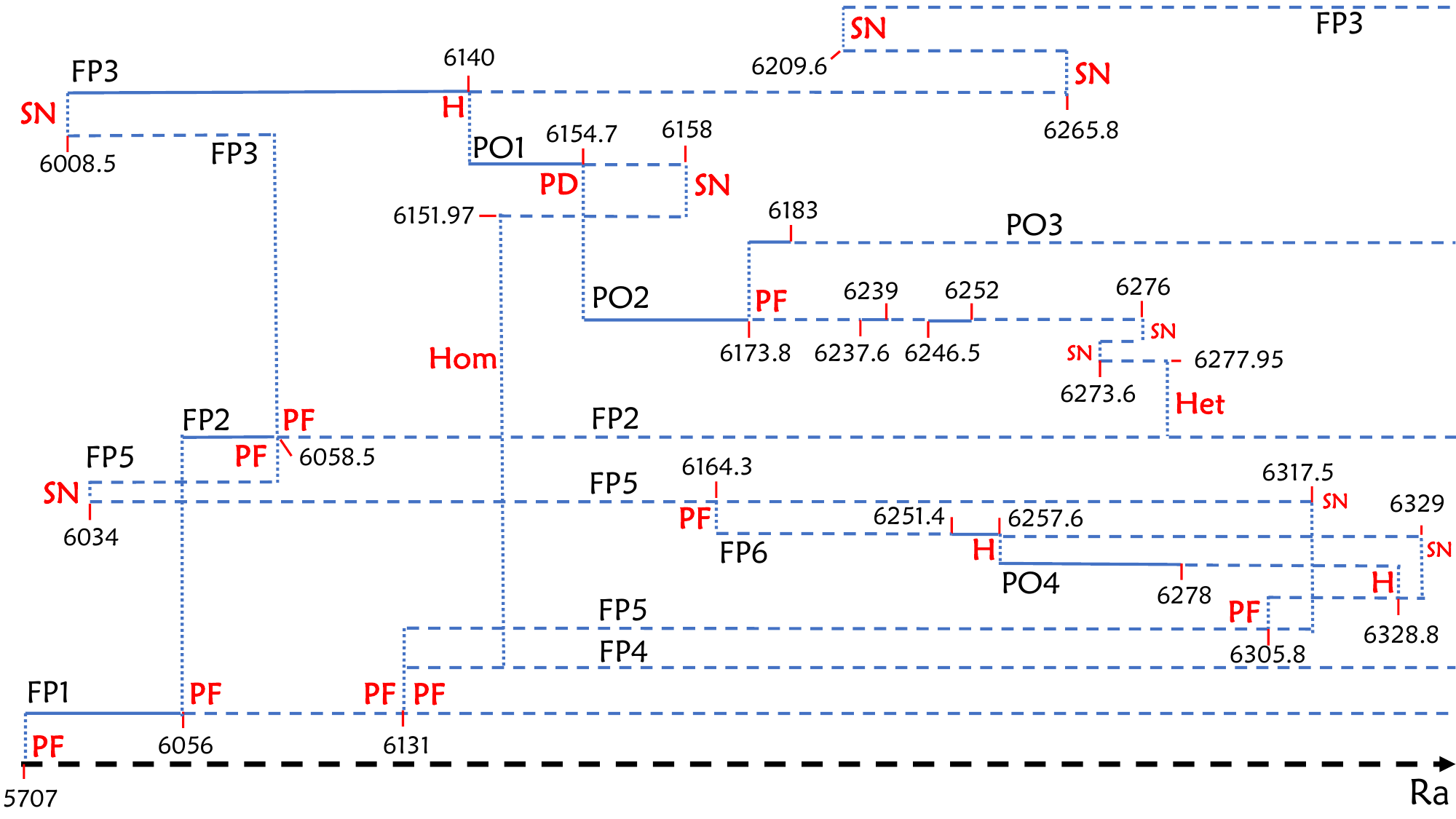}
    \captionsetup{font={footnotesize}}
    \captionsetup{width=13.5cm}
    \captionsetup{format=plain, justification=justified}
    \caption{\label{Schematic-BD} Schematic bifurcation diagram summarizing the origin and stability of all of the fixed points (FPs) and periodic orbits (POs) that we identified in the computational domain $[L_x, L_y, L_z] = [1, 8, 9]$. PF, SN, PD, H, Het and Hom are abbreviations for pitchfork, saddle-node, period-doubling, Hopf, heteroclinic and homoclinic bifurcations. The dotted vertical lines together with the solid red lines and numbers mark the Rayleigh numbers at which bifurcations occur. Solid and dashed horizontal lines signify stable and unstable states respectively.} 
\end{figure}

\par We have numerically investigated vertical thermal convection in the domain $[L_x, L_y, L_z] = [1, 8, 9]$, the configuration studied by \cite{Gao2018}, for Rayleigh number up to $Ra\approx6400$. In this Rayleigh number range, the system exhibits various spatio-temporally organized flow patterns and weak turbulence. Using the computational power of parallelized numerical continuation based on matrix-free Newton methods, we have computed invariant solutions, more specifically fixed points, periodic orbits, and homoclinic and heteroclinic orbits. 

\par We have situated all known solutions in the context of a bifurcation diagram. The diagrams shown in figures \ref{FP918} and \ref{largedomain_BDall} are presented in schematic form in figure \ref{Schematic-BD}. This diagram contains the names of the states and the bifurcations between them, along with their precise thresholds, and emphasizes the complexity of the bifurcation scenario. As was the case for \cite{Zheng2023part1}, all of the solution branches that we have found here are connected directly or indirectly to the laminar branch. This is not always so: our ongoing investigation has revealed branches which arise via saddle-node bifurcations and seem to be unconnected to the laminar state; see also figure 3 of \cite{Reetz2020b}. 

\par Compared to the narrow domain $[1, 1, 10]$ presented in \citet{Zheng2023part1}, the critical Rayleigh number for the primary instability of four spanwise-independent co-rotating rolls (called FP1 in both papers) in the spanwise-extended domain is only slightly lower. This is due  to the slight reduction in the vertical length from $L_z=10$ to $L_z=9$ or equivalently from $\lambda=2.5$ to $\lambda=2.25$ in the primary roll wavelength. However, secondary and tertiary branches exist at much lower Rayleigh numbers for the $[1, 8, 9]$ domain than for the $[1,1,10]$ domain, since the larger domain accommodates a wider variety of spanwise-varying patterns. 

\par We observe complicated bifurcation scenarios involving both spatial and temporal aspects. Spatially, parametric continuation reveals two types of branches. One set of branches consists of states which are aligned with the periodic directions $y$ and $z$: FP1 (2D rolls), FP2 (diamond rolls), FP5 (mustache rolls) and the closely related FP6. The other set of branches consists of states which are oriented diagonally: FP3 (thinning rolls) or the similar FP4. We observed two instances of simultaneous bifurcation to branches of states with different symmetries. We were able to explain this otherwise non-generic phenomenon as the breaking of $D_4$ symmetry of the parent branches FP1 and FP2. (In this highly symmetric geometry, $D_4$ symmetry is a subgroup of the full symmetry group of FP1 and FP2.) We confirmed this by computing and comparing the eigenvectors responsible for the simultaneous bifurcations.

\par Temporally, by following certain periodic orbit branches, PO1 and PO2, far from their onset via Hopf and period-doubling bifurcations, we have identified homoclinic and heteroclinic bifurcations that terminate these periodic-orbit branches. The fixed points at which these orbits spend an increasingly long time are aligned with the $y$, $z$ axes (FP2), or nearly so (FP4), while the excursions are to diagonal states. Thus, these periodic orbits and global bifurcations can also be seen as a manifestation of competition between aligned and diagonal states. Although this is well understood from a mathematical group-theoretic viewpoint, there may exist some physical or phenomenological interpretation of when and why aligned or diagonal states are favored. Another type of competition that we observe is between wavelengths: the heteroclinic orbit from FP2 can be interpreted as resulting from competition or interaction between states with wavenumbers 1 ($y$-wavelength $L_y$) and 2 ($y$-wavelength $L_y/2$). Indeed, the 1:2 mode interaction is a classic scenario leading to a robust heteroclinic orbit \citep{Armbruster1988, Mercader2002, Nore2003mode, Reetz2020a}.

\par The highest Rayleigh number that we have studied is $Ra=6407$, $12.3\%$ above the onset of convection (FP1 at $Ra=5707$). Even in this relatively small range of $Ra$, we have found a large variety of branches and bifurcation scenarios and there are certainly more to be discovered and analyzed. In particular, primary bifurcations from the base state can lead to secondary states containing spanwise-independent (or 2D) co-rotating rolls of many other wavelengths (all unstable at onset). These 2D rolls can also undergo secondary, tertiary and global bifurcations. The increasing number of branches with Rayleigh or Reynolds number is a general feature of the Navier-Stokes and Boussinesq equations. However, branches can also disappear by the same types of local bifurcations that create them, and periodic orbits can be destroyed by global bifurcations, both of which occur in our configuration. 

\par It has been conjectured that trajectories in chaotic and turbulent flows spend a substantial amount of time visiting unstable periodic orbits that are linked via their stable and unstable manifolds \citep{Cvitanovic1991, Kawahara2001, Suri2020, Crowley2022}. Computing unstable periodic orbits and understanding the bifurcations which produce and link them are thus relevant to better understanding and statistical measures of turbulent flows \citep{Clever1995, Graham2021}. In particular, reconstructing turbulence statistics using periodic orbits was explored by \citet{Chandler2013}, using around 50 periodic orbits embedded in turbulent two-dimensional Kolmogorov flow; see also \citet{Cvitanovic2013}. Extending this approach to three-dimensional turbulent thermal convection is one of the objectives of our future research.

\backsection[Acknowledgements]{We thank Sajjad Azimi, Florian Reetz and Omid Ashtari for fruitful discussions. We are grateful to Dwight Barkley, Philippe Beltrame, Pierre Gaspard, and Edgar Knobloch for their insights on heteroclinic cycles.}

\backsection[Funding]{This work was supported by the European Research Council (ERC) under the European Union's Horizon 2020 research and innovation programme (grant no. 865677).}

\backsection[Declaration of interests]{The authors report no conflict of interest.}

\backsection[Author ORCID]{
\newline
Zheng Zheng \href{https://orcid.org/0000-0002-9833-1347}{https://orcid.org/0000-0002-9833-1347};
\newline
Laurette S. Tuckerman \href{https://orcid.org/0000-0001-5893-9238}{https://orcid.org/0000-0001-5893-9238}; 
\newline
Tobias M. Schneider \href{https://orcid.org/0000-0002-8617-8998}{https://orcid.org/0000-0002-8617-8998}.}

\bibliographystyle{jfm}

\end{document}